\documentclass[twocolumn]{aastex631}
\usepackage{epsf}
\usepackage{graphics,graphicx}
\usepackage{color}
\usepackage{amssymb}
\usepackage{xcolor}
\usepackage{amsmath,amssymb}
\usepackage{longtable}
\usepackage{hyperref}
\hypersetup{colorlinks,citecolor=blue}
\def\~{$\sim$}

\shorttitle{FAST H{\sc i} and OH absorption line survey towards radio  AGN}
\shortauthors{Chandola et al.}
\begin{document}

\title{FAST survey of H{\sc i} and OH absorption towards  extragalactic radio sources}

\correspondingauthor{Yogesh Chandola}
\email{yogesh.chandola@pmo.ac.cn}

\author[0000-0002-3698-3294]{Yogesh Chandola}
\affiliation{Purple Mountain Observatory, Chinese Academy of Sciences,
10, Yuan Hua Road, Qixia District,
Nanjing, 210023, China}

\author[0000-0002-4464-8023]{D.J.Saikia}
\affiliation{Inter-University Centre for Astronomy and Astrophysics (IUCAA),
Pune University Campus, Ganeshkhind,
Pune, 411007, India
}

\author{Yin-Zhe Ma}
\affiliation{Department of Physics, Stellenbosch University, Matieland 7602, South Africa 
}
\author{Zheng Zheng}
\affiliation{National Astronomical Observatories, Chinese Academy of Sciences, 20A Datun Rd, Beijing, 100101, China}
\affiliation{Key Laboratory of Radio Astronomy and Technology, Chinese Academy of Sciences, Beijing, 100101, China}
\author[0000-0002-9390-9672]{Chao-Wei Tsai}
\affiliation{National Astronomical Observatories, Chinese Academy of Sciences, 20A Datun Rd, Beijing, 100101, China}
\affiliation{Institute for Frontiers in Astronomy and Astrophysics, Beijing Normal University, Beijing 102206, China}
\affiliation{School of Astronomy and Space Science, University of Chinese Academy of Sciences, Beijing 100049,
China}
\author{Di Li}
\affiliation{National Astronomical Observatories, Chinese Academy of Sciences, 20A Datun Rd, Beijing, 100101, China}

\author{Denis Tramonte}
\affiliation{Department of Physics, Xi'an Jiaotong-Liverpool University, 111 Ren'ai Road, Suzhou Dushu Lake Science and Education Innovation District, Suzhou Industrial Park, Suzhou 215123, P.R. China 
}

\author{Hengxing Pan}
\affiliation{Astrophysics, University of Oxford, Denys Wilkinson Buiding, Keble Road, Oxford OX1 3RH, UK 
}



\begin{abstract}
Neutral atomic hydrogen and molecular gas in the host galaxies of radio active galactic nuclei (AGN) can be traced using H{\sc i} 21-cm and OH-1667 MHz absorption lines to understand the fueling and feedback processes.  We present the results of an H{\sc i} and OH absorption survey with the Five-hundred-meter Aperture Spherical radio Telescope (FAST) towards 40 radio sources of low-intermediate radio luminosity ($\sim$10$^{23}$-10$^{26}$ W Hz$^{-1}$ at 1.4 GHz), red mid-infrared color (W2[4.6 $\mu$m]$-$W3[12 $\mu$m]  $>$ 2.5 mag) and redshift up to 0.35. From 13 sources with good data at H{\sc i} observing frequencies, we report the detection of H{\sc i} absorption towards 8 sources, 5 of which are new detections including 4 in the redshift range 0.25 to 0.35. Our detection rates are consistent with our previous results with dependence on the star-formation history of the host galaxy reflected in the mid-infrared  \textit{WISE} W2$-$W3 colors and the compactness of the radio source. We find no significant dependence of detection rates on radio luminosity or redshift.  We also find that H{\sc i} column densities are anti-correlated with the low-frequency spectral indices 
($\alpha_{\rm 150 MHz}^{\rm 1.4 GHz}$, $S_{\nu}\propto \nu^{-\alpha}$).    
We do not have any detection from 23 sources with good data at OH observing frequencies. However, by stacking the spectra we estimate the 3$\sigma$ upper limit of OH column density to be 2.27$\times$10$^{14}$$T_{\rm ex}$/10 K $\times$1/$f_{\rm c}$ cm$^{-2}$. By stacking the OH spectra for 7 associated H{\sc i} absorbers, we get a 3$\sigma$ upper limit of 3.47$\times$10$^{14}$ $T_{\rm ex}$/10 K $\times$1/$f_{\rm c}$ cm$^{-2}$ on OH column density and 1.78$\times$10$^{-7}$ on [OH]/[H{\sc i}] ratio. 
\end{abstract}



\section{Introduction} \label{sec:intro}
 Radio active galactic nuclei (AGN) are classified based on their optical emission line strengths into the high-excitation and low-excitation mode AGN, namely the high-excitation radio galaxies (HERGs) and low-excitation radio galaxies (LERGs) \citep{hine1979MNRAS.188..111H,laing1994ASPC...54..201L,buttiglione2010A&A...509A...6B,best2012MNRAS.421.1569B,heckman2014ARA&A..52..589H}. This classification is mainly attributed to the differences in accretion modes of the two types of AGN \citep{heckman2014ARA&A..52..589H}. HERGs are efficiently accreting mode radio  AGN with high Eddington ratios while LERGs are inefficiently accreting mode radio  AGN with low Eddington ratios \citep{heckman2014ARA&A..52..589H}. Fueling and feedback processes are also believed to be different in the two types of AGN. LERGs are said to be fueled due to the cooling of hot halo gas through Bondi accretion \citep{bondi1952MNRAS.112..195B,allen2006MNRAS.372...21A,hardcastle2007MNRAS.376.1849H} or chaotic accretion \citep{gaspari2013MNRAS.432.3401G, tremblay2018ApJ...865...13T}  while HERGs are said to be fueled due to the mergers \citep{hopkins2006ApJS..163....1H,ellison2013MNRAS.435.3627E} or secular processes \citep{kormendy2013ARA&A..51..511K}. Also, the feedback in LERGs is largely in the form of radio jets, while it is mainly due to radiation in HERGs \citep{heckman2014ARA&A..52..589H}.  AGN feedback can have significant positive or negative effects on star formation in the host galaxy \citep{harrison2017NatAs...1E.165H, mulcahey2022A&A...665A.144M,davies2024MNRAS.528.4976D, harrison2024Galax..12...17H}. Mechanical feedback due to jet-interstellar medium (ISM) interaction can cause gaseous outflows \citep{morganti1998AJ....115..915M,morganti2003ApJ...593L..69M, morganti2013Sci...341.1082M}. However, it also depends upon how the jet couples with the ISM. Simulations suggest that coupling efficiency depends on radio power and jet orientation \citep{mukherjee2018MNRAS.479.5544M,mukherjee2018MNRAS.476...80M}. Low-power highly inclined jets towards the disk can get trapped in the ISM and have a longer interaction time than higher-power and less inclined jets \citep{mukherjee2018MNRAS.479.5544M,mukherjee2018MNRAS.476...80M}. Jet-cloud interactions may also significantly affect the evolution of the radio source itself \citep[see][for a review]{2022JApA...43...97S}. While sources with strong radio power could evolve into large radio sources, some weak radio sources can be trapped in the dense interstellar medium and could die early before they evolve into larger radio sources \citep{kunertbajraszewska2010MNRAS.408.2261K, antao2012ApJ...760...77A}. 
  At the lower redshifts (z$<$0.1) and low radio luminosities, it is the LERGs which dominate the radio source population \citep{heckman2014ARA&A..52..589H}.
 
  Cold gas plays an important role in star formation and, hence in the evolution of galaxies. Hydrogen being the most abundant element in the Universe largely forms the gas reservoir for star-formation activity and fueling the AGN. Studying atomic hydrogen (H{\sc i}) in the host galaxies of radio  AGN can provide answers to the questions related to fueling the central engine and its feedback to the host galaxy and nearby environment. Hence it is important to study the kinematics and distributions of H{\sc i}  gas in the host galaxies of radio  AGN and their surroundings to understand their co-evolution. In the literature, 21-cm H{\sc i} absorption towards radio AGN has been used to probe the gas properties in the host galaxies and circumnuclear regions of radio  AGN \citep[e.g.][]{vermeulen2003A&A...404..861V, gupta2006MNRAS.373..972G, chandola2011MNRAS.418.1787C, chandola2013MNRAS.429.2380C,chandola2020MNRAS.494.5161C, gereb2015A&A...575A..44G, maccagni2017A&A...604A..43M, grasha2019ApJS..245....3G, murthy2021A&A...654A..94M,murthy2022A&A...659A.185M, morganti2023A&A...678A..42M}. Unlike H{\sc i} emission, H{\sc i} absorption is detected from the gas clouds intersecting   
  lines of sight to the background radio source. Hence the detection also depends upon geometrical as well as physical factors. However, due to its dependence on the brightness of the background source H{\sc i} absorption is advantageous to detect the neutral hydrogen at higher redshifts and smaller scales \citep[see][for a review]{morganti2018A&ARv..26....4M}. The compact radio AGN (projected linear sizes less than a few kiloparsecs) have higher H{\sc i} absorption detection rates compared to extended large radio sources (projected linear sizes $>$ 20 kiloparsecs), mainly because these sources have larger covering factors and trace the denser regions of H{\sc i} gas \citep{pihlstrom2003A&A...404..871P,vermeulen2003A&A...404..861V,gupta2006MNRAS.373..972G,chandola2011MNRAS.418.1787C,curran2013MNRAS.431.3408C}. Also, the host galaxies of compact radio AGN have been suggested to be gas-rich as compared to the host galaxies of larger radio sources at nearby redshifts \citep{emonts2010MNRAS.406..987E}.  In our earlier work,  we found that in the low-intermediate radio luminosity samples, LERGs have lower H{\sc i} absorption detection rates compared to HERGs largely because of their gas and dust-poor host galaxies \citep{chandola2020MNRAS.494.5161C}. H{\sc i} absorption detection rates largely depend upon the mid-infrared WISE W2$-$W3 colors of the host galaxies and the compactness of the radio sources at relatively nearby redshifts \citep{chandola2017MNRAS.465..997C,chandola2020MNRAS.494.5161C}. At higher redshifts, the detection rates have been affected by selection effects such as by samples of higher UV and radio luminosities in earlier observations. This has resulted in largely low detection rates at higher redshifts either due to high ionisation or high spin temperature or both \citep{curran2008MNRAS.391..765C, aditya2016MNRAS.455.4000A, aditya2018MNRAS.481.1578A,curran2019MNRAS.484.1182C,curran2024PASA...41....7C}.

\subsection{Low-luminosity radio sources, H{\sc i} absorption and FAST}
Due to the sensitivity limitations of radio telescopes, the early H{\sc i} absorption experiments were largely done towards bright radio sources selected from flux density-limited samples. This has biased the H{\sc i} absorption samples towards the higher radio luminosity sources ($>$10$^{26}$ W Hz$^{-1}$ at 1.4 GHz), especially at higher redshifts \citep[e.g.][]{aditya2016MNRAS.455.4000A, aditya2018MNRAS.473...59A,aditya2018MNRAS.481.1578A}. Since the low radio luminosity radio sources follow a different evolutionary trajectory/phase compared to higher luminosity sources, it is important to study the H{\sc i} environment and its evolution for these sources. In the past more than one decade, efforts towards studying H{\sc i} absorption towards low-intermediate radio luminosity ($\sim$10$^{23}$-10$^{26}$ W Hz$^{-1}$ at 1.4 GHz) sources have increased though limited to nearby radio sources \citep{chandola2020MNRAS.494.5161C,maccagni2017A&A...604A..43M}. \cite{maccagni2017A&A...604A..43M} studied H{\sc i} absorption towards approximately 250 sources at low redshifts ($z <$ 0.25) and low-intermediate radio luminosities.  The 19-beam receiver of the Five-hundred-meter Aperture Spherical radio Telescope \citep[FAST,][]{Nan2011IJMPD..20..989N} allows probing the H{\sc i} gas in lower radio luminosity sources to slightly larger redshifts of up to 0.35 due to its wide bandwidth and better sensitivity. Thus it helps in understanding the redshift evolution of the H{\sc i} gas environment of low radio luminosity sources, and also making comparisons with higher luminosity objects in limited redshift range to understand the effects of radio luminosity on H{\sc i} properties. FAST with its unprecedented sensitivity and broad bandwidth may also help to detect broad blue-shifted profiles reflecting outflowing gas due to jet-cloud interactions or radiation or starburst \citep{su2023ApJ...956L..28S}. In this paper, we present results from the survey of H{\sc i} absorption towards the radio AGN of low radio luminosities up to a redshift of 0.35. In the redshift range 0.25-0.35 this is the lowest radio luminosity survey than previous surveys. In this paper, we report the H{\sc i} absorption detection towards the four lowest radio luminosity sources to date in this redshift range.

\begin{table*}
	\caption{Radio sources observed with the FAST.}
	\begin{center}
		\scriptsize{ 
			\begin{tabular}{l l l l l l l l l  }
				\hline
				(1) &   (2)& (3) & (4)& (5) & (6) &(7)&(8) & (9) \\
				Source name &   Redshift & LERG/HERG & S$_{\rm NVSS}$ & S$_{\rm FIRST}$ & S$_{\rm FIRST}$ (peak) & $\alpha_{\rm 150 MHz}^{\rm 1.4 GHz}$ & log $P_{\rm 1.4 GHz}$& W2$-$W3 \\
&   & &mJy &mJy & mJy beam$^{-1}$ &   & W Hz$^{-1}$& mag \\
\hline
SDSS J022246.94$-$093848.7 &  0.30617$\pm$0.00004 & LERG & 56.7 & 57.38 & 55.32 & 0.19 &25.1&4.05 \\
SDSS J081437.98+172208.3 &   0.17919$\pm$0.00004 & LERG & 37.8 & 53.92 & 51.86 & $< -$0.15\tablenotemark{\scriptsize{a}} &24.4&2.60 \\
SDSS J081755.21+312827.4 &   0.12376$\pm$0.00002 & HERG & 69.2 & 61.48 & 53.3 & 0.63 &24.4&3.81 \\
SDSS J083216.03+183212.1 &   0.15411$\pm$0.00002 & HERG & 896.0 & 874.24 & 852.06 & $-$0.41 &25.7&2.77 \\ 
SDSS J083548.14+151717.0 &   0.16838$\pm$0.00003 & LERG & 44.8 & 48.66 & 46.31 &  &24.5&2.57 \\
SDSS J090410.36+024744.8 &   0.27691$\pm$0.00003 & LERG & 45.2 & 48.26 & 46.94 & $< -$0.47\tablenotemark{\scriptsize{a}} &24.9&3.68 \\
SDSS J092527.55+072641.6 &   0.12896$\pm$0.00002 & LERG & 104.4 & 103.07 & 97.06 &  0.56&24.6&3.61 \\
SDSS J092924.92+193421.0 &   0.30330$\pm$0.00007 & LERG & 112.3 & 111.27 & 106.05 & $-$0.55 &25.3&3.90 \\
SDSS J093242.81$-$003948.8 &   0.23534$\pm$0.00001 & HERG & 44.2 & 46.16 & 45.32 & $< -$0.51\tablenotemark{\scriptsize{a}} &24.7&3.31 \\
SDSS J094310.82+295203.6 &   0.29941$\pm$0.00006 & LERG & 54.4 & 57.68 & 55.66 & 0.12 &25.1& $>$2.67 \\
SDSS J095058.69+375758.8 &   0.04053$\pm$0.00001 & HERG & 66.6 & 68.26 & 67.51 & $< -$0.54\tablenotemark{\scriptsize{a}} &23.4&3.01 \\
SDSS J102453.63+233234.0 &   0.16462$\pm$0.00002 & LERG & 108.9 & 125.13 & 121.33 & $-$0.26&24.8&2.77 \\
SDSS J110701.20+182548.8 &   0.17856$\pm$0.00002 & HERG & 162.7 & 159.93 & 142.03 & 0.77 &25.1&2.79 \\
SDSS J114538.51+442021.9 &   0.29974$\pm$0.00004 & LERG & 362.3 & 327.31 & 309.35 & 0.46&26.0 & 2.78\\
SDSS J115712.38$-$032107.7 &   0.08202$\pm$0.00002 & LERG & 50.7 & 52.71 & 50.61 & 0.29 &23.9& 2.84\\
SDSS J121755.30$-$033723.3 &   0.18229$\pm$0.00003 & HERG & 213.6 & 208.26 & 176.68 & 0.63&25.3& 2.57 \\ 
SDSS J122113.25$-$024859.5 &   0.11300$\pm$0.00002 & LERG & 100.3 & 105.56 & 98.16 & $-$0.29&24.5& 2.57 \\
SDSS J122228.47+171437.3 &   0.31894$\pm$0.00006 & HERG & 49.6 & 47.4 & 47.17 & 0.14 &25.1&2.59 \\
SDSS J124419.96+405136.8 &   0.24934$\pm$0.00002 & HERG & 361.2 & 367.63 & 345.89 & 0.59 &25.8 &2.68  \\
SDSS J124707.32+490017.9 &  0.20691$\pm$0.00002 & LERG & 1204.8 & 1212.7 & 1037.73 & 0.40&26.1&2.54 \\
SDSS J132522.00+035848.9 &   0.25479$\pm$0.00005 & LERG & 113.5 & 113.29 & 111.12 & 0.34&25.3 & 2.56\\
SDSS J132859.25+173842.3 &   0.18035$\pm$0.00003 & HERG & 158.6 & 158.78 & 152.77 & 0.30&25.1 & 2.92 \\
SDSS J133242.53+134253.8 &   0.28287$\pm$0.00002 & HERG & 57.1 & 52.67 & 47.74 & 0.81&25.1& 2.98 \\
SDSS J135223.46$-$015648.4 &   0.16694$\pm$0.00001 & HERG & 529.6 & 552.19 & 479.26 & 0.72 &25.6& 2.53 \\
SDSS J141327.22+550529.2 &   0.28156$\pm$0.00006 & LERG & 125.9 & 128.48 & 125.77 & $-$0.039&25.4& 2.96 \\
SDSS J143806.13+190954.9 &   0.18854$\pm$0.00002 & HERG & 45.4 & 44.52 & 43.16 & 0.76 &24.6& 2.84 \\
SDSS J144920.71+422101.2 &   0.17862$\pm$0.00006 & HERG & 155.7 & 165.64 & 159.08 & $-$0.38 &25.0& 3.20 \\
SDSS J145844.79+372021.5 &  0.33331$\pm$0.00008 & LERG & 214.8 & 269.97 & 266.39 & $-$0.31 &25.7& 2.91 \\
SDSS J152142.58+181438.2 &   0.15043$\pm$0.00006 & HERG & 92.1 & 93.27 & 90.9 & 0.74&24.7 & 3.86\\
SDSS J153016.25+375831.2 &   0.15171$\pm$0.00002 & HERG & 104.9 & 99.82 & 98.53 &$< -$0.62\tablenotemark{\scriptsize{a}} &24.7 & 3.27 \\
SDSS J153229.40+015133.7 &   0.12326$\pm$0.00001 & LERG & 79.0 & 81.92 & 79.53 & 0.42&24.5& 3.37 \\
SDSS J153836.11+552541.4 &   0.19117$\pm$0.00005 & HERG & 209.2 & 209.84 & 184.22 & 0.72&25.3 & 2.60 \\
SDSS J154345.80+110935.9 &   0.10231$\pm$0.00002 & LERG & 48.1 & 48.22 & 47.73 & 0.21&24.1 & 2.70 \\
SDSS J155903.43+230828.7 &   0.19318$\pm$0.00004 & LERG & 43.6 & 44.95 & 43.67 & 0.29&24.6 & 2.53\\
SDSS J155927.67+533054.4 &   0.17921$\pm$0.00001 & HERG & 182.7 & 182.35 & 170.43 &0.97 &25.2 & 2.86 \\
SDSS J162033.43+173955.5 &   0.16361$\pm$0.00002 & LERG & 107.1 & 111.44 & 104.23 & $< -$0.26\tablenotemark{\scriptsize{a}} &24.8 & 3.27\\
SDSS J213333.31$-$071249.2 &   0.08654$\pm$0.00001 & HERG & 189.8 & 200.5 & 194.78 &0.57 &24.5 & 4.26 \\
SDSS J230551.18$-$104052.2 &   0.18970$\pm$0.00002 & LERG & 69.9 & 70.34 & 66.42 &0.21 &24.8 & 3.36\\
SDSS J233515.92$-$011216.8 &   0.26988$\pm$0.00003 & LERG & 45.7 & 45.85 & 42.07 &0.72 &25.0 & 3.85\\
SDSS J235400.91$-$003449.5 &   0.32593$\pm$0.00008 & LERG & 66.6 & 67.72 & 66.43 &$< -$0.31\tablenotemark{\scriptsize{a}} &25.2& 2.54 \\
				\hline
			\end{tabular}
		}
	\end{center}
	\vspace{-0.4 cm}
	\begin{flushleft}
 \tablecomments{
		Column 1: Source name; 
		Column 2; Heliocentric redshifts from Sloan Digital Sky Survey \citep[SDSS,][]{ahumada2020ApJS..249....3A} DR16;
		Column 3: LERG and HERG classification based on \cite{best2012MNRAS.421.1569B};
		Column 4: NVSS flux density in the units of mJy;
		Column 5: FIRST integrated flux density in units of mJy;
		Column 6: Peak FIRST flux density in the units of mJy/beam;
       Column 7: Spectral index ($\alpha_{\rm \rm 150 MHz}^{\rm 1.4 GHz}$) from \cite{degasparin2018MNRAS.474.5008D} estimated using flux density values from  TIFR-GMRT Sky Survey \citep[TGSS;][]{intema2017A&A...598A..78I} and NVSS. Spectral index is defined as $S_{\nu} \propto \nu^{-\alpha}$;
  Column 8: logarithm of luminosity at 1.4 GHz in units of W Hz$^{-1}$ estimated using NVSS flux densities; Column 9: AllWISE W2[4.6 $\mu$m]$-$W3[12 $\mu$m] color in Vega magnitude.\\
   \tablenotetext{a}{Source is not detected in TGSS. The spectral index upper limit has been estimated using 3$\sigma$ upper limit on noise value in TGSS image.}}
	\end{flushleft}
	\label{sample}
\end{table*}

\subsection{OH absorption and FAST}
Tracing the cold molecular gas which is directly associated with star-formation activity in the
host galaxies is also important to understand the feedback effects of AGN. Diffuse molecular
gas in the interstellar medium can be also traced with the hydroxyl molecule (OH). OH can be
detected in four hyperfine transitions, mostly at 1667 MHz. Other lines at 1665 MHz, 1612 MHz
and 1720 MHz are weak or absent \citep{mcbride2013ApJ...774...35M}. To date, only two radio AGN, B3 1504+377 and PKS
1413+135, have been detected with associated OH absorption lines \citep{kanekar2002A&A...381L..73K}. Using FAST for the pilot OH
absorption study towards 3 radio AGN detected in H{\sc i} absorption, \cite{zheng2020MNRAS.499.3085Z} obtained an
upper limit on the OH column density ($\sim$1.57$\times$10$^{14}$ $T_{\rm ex}$/10 K $\times$1/$f_{\rm c}$   cm$^{-2}$ ) and its ratio with
H{\sc i} column density ($<$ 5.45$\times$10$^{-8}$ ).
Cosmic evolution of several physical constants such as the electron-proton mass ratio $\rm \mu = m_p/m_e$, the fine structure constant $\alpha = e^{2}/\hbar$, and the proton g-factor $g_p$ can also be constrained using the cospatial H{\sc i} and OH absorption lines \citep{kanekar2005PhRvL..95z1301K}. Large
surveys to detect the rare coincidence of H{\sc i} 21 cm and OH 18 cm absorption lines can be
conducted with the large bandwidth of the FAST 19-beam receiver.  Many sources
within the redshift of 0.1499-0.35 can be searched for simultaneous H{\sc i} absorption as well as OH absorption using the FAST 19-beam receiver.

In this paper, we present the results of H{\sc i} and OH absorption survey using FAST  towards 40 low-intermediate luminosity ($\sim$10$^{23}$-10$^{26}$ W Hz$^{-1}$ at 1.4 GHz) radio sources from \cite{best2012MNRAS.421.1569B} with red mid-IR colour (W2$-$W3 $>$ 2.5 mag) up to the redshift of 0.35. The paper is arranged as follows. In Section 2, we describe our sample selection, observations and data reduction. In Section 3, we present our results and in Section 3.1, we list individual H{\sc i} absorption cases. We discuss our results in Section 4 and finally, we summarise our results in Section 5. Throughout the paper, we have used concordance cosmology $\Omega_{\rm m}=$0.3, $\Omega_{\Lambda}=$0.7 and $H_{\rm 0}=$70 km s$^{-1}$ Mpc$^{-1}$.

\begin{table*}
	\caption{Observational details of the search for associated H{\sc i} and OH absorption.}
	\begin{center}
		\scriptsize{
			\begin{tabular}{ l l c l c l c c  c }
				
				\hline
	(1)& (2)& (3)& (4)& (5)& (6)& (7) \\
Source name & H{\sc i} freq.& Beam(s) used  & OH freq.& Beam(s) used & Observation     &     ON source    \\
             & (MHz)  	&	& (MHz)&  & date     & Time (s)  \\
				\hline
SDSS J022246.94$-$093848.7 & 1087.4586 & RFI  & 1276.5253& RFI  & 2022 Jan 29  			  & 540 \\
SDSS J081437.98+172208.3   & 1204.5606 & RFI  & 1413.9867& M01, M14  & 2021 Oct 02 			  & 540\\
SDSS J081755.21+312827.4   & 1263.9761 & RFI  & 1483.7323& OR  & 2021 Oct 05 			  & 360\\
SDSS J083216.03+183212.1   & 1230.7369 & RFI       & 1444.7141& M01       & 2021 Oct 20  			  & 180\\ 
SDSS J083548.14+151717.0   & 1215.7053 & RFI  & 1427.0691& M01, M14  & 2021 Oct 20  			  & 540\\
SDSS J090410.36+024744.8   & 1112.3773 & M01, M14  & 1305.7764& M01  & 2021 Oct 23  \& 2022 Jan 29     & 750\\
SDSS J092527.55+072641.6   & 1258.1542 & RFI  & 1476.8982& OR  &  2021 Oct 22  			  &  360 \\
SDSS J092924.92+193421.0   & 1089.8533 & RFI  & 1279.3363& RFI  & 2021 Oct 18  			  & 180 \\
SDSS J093242.81$-$003948.8 & 1149.8096 & RFI  & 1349.7167& M01, M14  & 2022 Jan 30   			  &  720 \\
SDSS J094310.82+295203.6   & 1093.1159 & M01       & 1282.1662& RFI       & 2021 Oct 23   			  & 540 \\
SDSS J095058.69+375758.8   & 1365.0791 & M01       & 1602.4132& OR       &  2021 Nov 10  			  & 360\\
SDSS J102453.63+233234.0   & 1219.6302 & RFI  & 1431.6764& M01, M14  & 2021 Oct 19  			  &  180\\
SDSS J110701.20+182548.8   & 1205.2044 & RFI  & 1414.7426& M01, M14  & 2021 Nov 11  			  & 180 \\
SDSS J114538.51+442021.9   & 1092.8384 & M01       & 1282.8404& RFI       & 2021 Nov 18  			  &180 \\ 
SDSS J115712.38$-$032107.7 & 1312.7352 & M01, M14  & 1540.9687& OR  & 2022 Jan 30   			  &  720 \\
SDSS J121755.30$-$033723.3 & 1201.4022 & RFI  & 1410.2792& M01, M14  & 2022 Feb 03   			  & 180 \\ 
SDSS J122113.25$-$024859.5 & 1276.1956 & RFI  & 1498.0764& OR  & 2021 Dec 02  			  & 180 \\
SDSS J122228.47+171437.3   & 1076.9298 & M01, M14  & 1264.1659& RFI  & 2022 Jan 30  			  & 720\\
SDSS J124419.96+405136.8   & 1136.9249 & RFI       & 1334.5919& M01       & 2021 Nov 22  			  &180\\ 
SDSS J124707.32+490017.9   & 1176.8945 & RFI       & 1381.5106& M01       & 2021 Nov 23   			  &180 \\ 
SDSS J132522.00+035848.9   & 1131.9868 & RFI     & 1328.7952& M01, M14  & 2021 Dec 19   			  & 180\\
SDSS J132859.25+173842.3   & 1203.3767 & RFI     & 1412.5971& M01, M14  & 2021 Dec 17   			  & 180\\
SDSS J133242.53+134253.8   & 1107.2094 & M01, M14  & 1299.7100& M01, M14  & 2022 Feb 02  			  & 615 \\
SDSS J135223.46$-$015648.4 & 1217.2055 & M01, M14  & 1428.8301& M01, M14  & 2021 Dec 12   			  & 180 \\
SDSS J141327.22+550529.2   & 1108.3412 & Failed          & 1301.0386& Failed  & 2021 Dec 20 			  & 180\\
SDSS J143806.13+190954.9   & 1195.0845 & M01        & 1402.8632& M01        & 2022 Jan 31  			  & 900\\
SDSS J144920.71+422101.2   & 1205.1431 & RFI        & 1414.6706& M01   & 2021 Dec 12 			  & 180\\
SDSS J145844.79+372021.5   & 1065.3230 & M01   & 1250.5411 & RFI   &  2022 Jan 03  			  & 180\\
SDSS J152142.58+181438.2   & 1234.6738 & RFI        & 1449.3355& OR   &  2022 Jan 03 			  &180\\ 
SDSS J153016.25+375831.2   & 1233.3016 & RFI  	     & 1447.7247 & M01  	     & 2021 Dec 20  			  & 180\\
SDSS J153229.40+015133.7   & 1264.5387 & RFI       & 1484.3928& OR   &  2022 Jan 04  			  & 180\\
SDSS J153836.11+552541.4   & 1192.4459 & RFI        & 1399.7658& M01        & 2022 Jan 07  			  &180\\ 
SDSS J154345.80+110935.9   & 1288.5720 & RFI   & 1512.6044& OR   & 2022 Jan 04  			  &  540\\
SDSS J155903.43+230828.7   & 1190.4371 & RFI   & 1397.4078& M01, M14   & 2022 Feb 01   			  & 720\\
SDSS J155927.67+533054.4   & 1204.5401 & RFI 	     & 1413.9627 & M01 	     & 2022 Jan 05  			  & 180 \\
SDSS J162033.43+173955.5   & 1220.6888 & RFI   & 1432.9191& M01, M14   & 2022 Jan 05  			  & 180\\
SDSS J213333.31$-$071249.2 & 1307.2742 & M14        & 1534.5583& OR        & 2021 Sep 20   			  & 180\\
SDSS J230551.18$-$104052.2 & 1193.9193 & RFI     & 1401.4953& M01, M14   & 2021 Sep 27   			  & 540 \\
SDSS J233515.92$-$011216.8 & 1118.5354 & M01, M14   & 1313.0052& M01, M14   & 2022 Jun 01   			  &  720 \\
SDSS J235400.91$-$003449.5 & 1071.2524 & M01        & 1257.5015& RFI        & 2021 Sep 28   			  &  360 \\

				\hline
			\end{tabular}
		}
	\end{center}
 \begin{flushleft}
 \tablecomments{
		Column 1: Source name; Column 2: H{\sc i} 21-cm redshifted line frequency in MHz; Column 3: Beam(s) used in final H{\sc i} spectra;
		Column 4: OH redshifted 1667.359 MHz line frequency in MHz; Column 5: Beam(s) used in final OH spectra;
		Column 6: Date of observation; 
		 Column 7: ON target source time per beam in seconds.\\
RFI: Radio Frequency Interference, OR: Out of frequency range, Failed: Observation failed due to technical problems.}
	\end{flushleft}
	\label{FASTobs}
\end{table*}

\section{Sample, observation and data reduction }
\label{sec:sec2}

\begin{figure}
    \centering
    \includegraphics[scale=0.55]{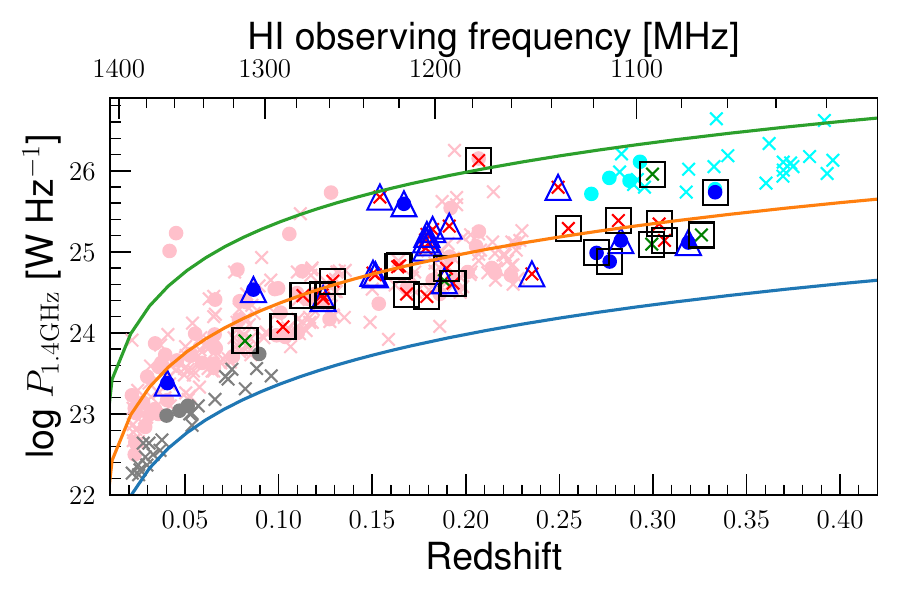}
    \caption{Radio luminosity at 1.4 GHz vs redshift. H{\sc i} observing frequencies are shown in the upper X-axis. Sources observed with FAST, shown with black squares and blue triangles represent LERGs and HERGs respectively. Sources with H{\sc i} absorption detections are marked with blue-filled circles while non-detections and RFI-affected or failed observations are marked with green and red crosses respectively. For comparison, we have also shown the sources with previous H{\sc i} absorption studies of low-intermediate radio luminosity below a redshift of 0.4 from \citet{maccagni2017A&A...604A..43M}, \citet{murthy2021A&A...654A..94M}, and
    \citet{yuqingzheng2023ApJ...952..144Y} with pink, cyan and grey colored circles (detections) and crosses (non-detections) respectively. Solid blue, orange and green lines mark the sources with flux densities of 10 mJy, 100 mJy and 1000 mJy at 1.4 GHz respectively.}
    \label{fig:lumz}
\end{figure}

We selected a sample of 76 radio sources from \citet{best2012MNRAS.421.1569B} with their peak flux densities $>$ 40 mJy beam$^{-1}$ at 1.4 GHz in the Faint Images of the Radio Sky at Twenty-Centimetres \citep[FIRST,][]{becker1995ApJ...450..559B} survey, mid-infrared AllWISE \citep{Cutri2014yCat.2328....0C} colour W2[4.6 $\mu$m]$-$W3[12 $\mu$m] $>$ 2.5 mag and redshift $<$ 0.35. The redshift cutoff of below 0.35 was given so that observing H{\sc i} frequency lies above 1050 MHz, the low-frequency limit of the FAST 19-beam feed receiver. We also selected sources with unresolved radio structures in the FIRST survey based on the ratios of the NRAO-VLA Sky Survey \citep[NVSS,][]{Condon1998AJ....115.1693C} flux density to the FIRST flux density ($<$1.2) and FIRST integrated to peak radio flux densities ($<$1.2). We further removed sources with earlier studies in H{\sc i} absorption except those with redshift  $>$ 0.1499 (corresponding to observing frequency of 1450 MHz for OH 1667 MHz spectral line) to study OH absorption. This left us with 55 sources. Further 9 sources were removed which had previous observations with FAST which gave us 46 sources. Of these 46 sources, 40 were observed with FAST. Of these 40, 22 have been classified as LERGs and 18 as HERGs by \cite{best2012MNRAS.421.1569B}. 
We have listed these 40 sources in Table~\ref{sample}. For 39 sources, we also have low-frequency spectral indices ($\alpha_{\rm 150 MHz}^{\rm 1.4 GHz}$, $S_{\nu} \propto$ $\nu^{-\alpha}$)
from \citet{degasparin2018MNRAS.474.5008D} estimated using flux densities from  TIFR-GMRT Sky Survey \citep[TGSS;][]{intema2017A&A...598A..78I} and NVSS. The 1.4 GHz radio luminosities of these sources are in the range $\sim$10$^{23.4}$ W Hz$^{-1}$ to 10$^{26.1}$ W Hz$^{-1}$ with a median value $\sim$10$^{25}$ W Hz$^{-1}$ (see Fig.~\ref{fig:lumz}). In Fig.~\ref{fig:lumz}, we have also shown the relative position of earlier H{\sc i} absorption studies towards low-intermediate radio luminosity sources from \citet{maccagni2017A&A...604A..43M},  \citet{murthy2021A&A...654A..94M} and \cite{yuqingzheng2023ApJ...952..144Y} in the luminosity-redshift plot for comparison. Most of the sources in our sample in the redshift range 0.25-0.35 have radio luminosities lower than those from \cite{murthy2021A&A...654A..94M}.

We observed these 40 radio sources with the FAST in ON-OFF mode from September 2021 to June 2022 (project ID: PT2021\_0034). The observational details are provided in Table~\ref{FASTobs}.  Of these 40 sources, for one source, SDSS J141327.22+550529.2, observations failed due to a problem in the control system. Of the remaining 39, 30 sources could also be observed for OH absorption simultaneously. We used an ON/OFF source period of 180 seconds in each ON-OFF cycle, except for J133242.53$+$134253.8 where we used 205 seconds. Following \citet{zheng2020MNRAS.499.3085Z}, the OFF source positions for beam M01 were selected $\sim$ 11.8\arcmin\hspace{1mm} towards the east in right ascension such that another beam M14 directs towards the ON source position. The data was sampled with a sampling time of $\sim$1 second. For calibration, we injected high noise ($\sim$10 K) for $\sim$1 second after every 10 seconds. 

The data were reduced using a Python-based spectral line data reduction pipeline developed by us. The pipeline first combines the data from all the raw FITS files and then separates it for each ON-OFF cycle. Then the pipeline converts the power into antenna temperature ($T_{\rm a}$) for each cycle using the equation,
\begin{equation}
    T_{\rm a} = \frac{P_{\rm caloff}T_{\rm cal}}{P_{\rm calon}-P_{\rm caloff}},
\end{equation}
where $P_{\rm calon}$ and $P_{\rm caloff}$ are power measured when noise injection is ON and OFF respectively. $T_{\rm cal}$ is the temperature of noise injected.  After conversion to $T_{\rm a}$, RFI affected time and frequency stamps are masked and average values of ON source antenna temperature ($T_{\rm a, ON}$) and OFF source antenna temperature  ($T_{\rm a, OFF}$) are obtained for each cycle. For each cycle, we obtain the brightness temperature $T_{\rm b}$ by subtracting $T_{\rm a, OFF}$ from $T_{\rm a, ON}$. The brightness temperature $T_{\rm b}$ spectra are obtained for different cycles and averaged together. A high-degree polynomial and the sine wave function are fitted outside the expected spectral line portion in the spectrum. These functions are subtracted from this spectrum for XX and YY polarisations separately to subtract the continuum and remove the ripples.  The spectrum is then averaged for XX and YY polarisations and converted to mJy using the gains for different beams from \cite{jiang2020RAA....20...64J} when the zenith angle is less than 26.4$^{\circ}$. For the cases, when the zenith angle is greater than 26.4$^{\circ}$ gains are estimated using the equation (3) of \cite{zhangkai2019SCPMA..6259506Z}. The frequencies are then converted to velocities using the optical definitions and corrected for Doppler shift due to the Earth's motion around the Sun using \texttt{astropy} library. Finally, the spectra from beams M01 and M14 were averaged together to get the final spectrum when both beams had good data. The typical spectral resolution of the final spectrum is $\sim$2 km s$^{-1}$ (7.63 kHz) which we further Hanning smoothed to the velocity resolution given in Tables \ref{hiabsresults} and \ref{ohabsresults}.

\section{Results}
\begin{figure*}[htb!]
    \centering
    \includegraphics[scale=0.50]{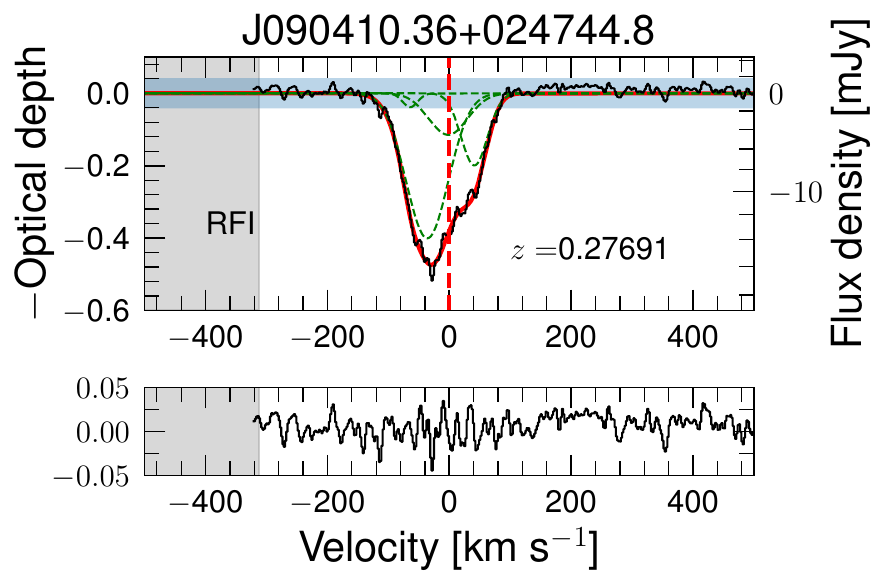}
    \includegraphics[scale=0.50]{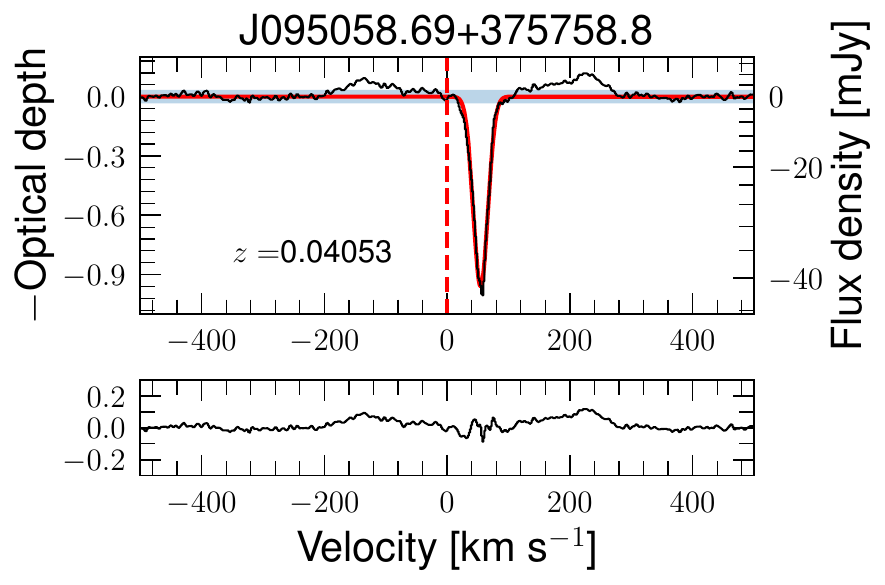}
    \includegraphics[scale=0.50]{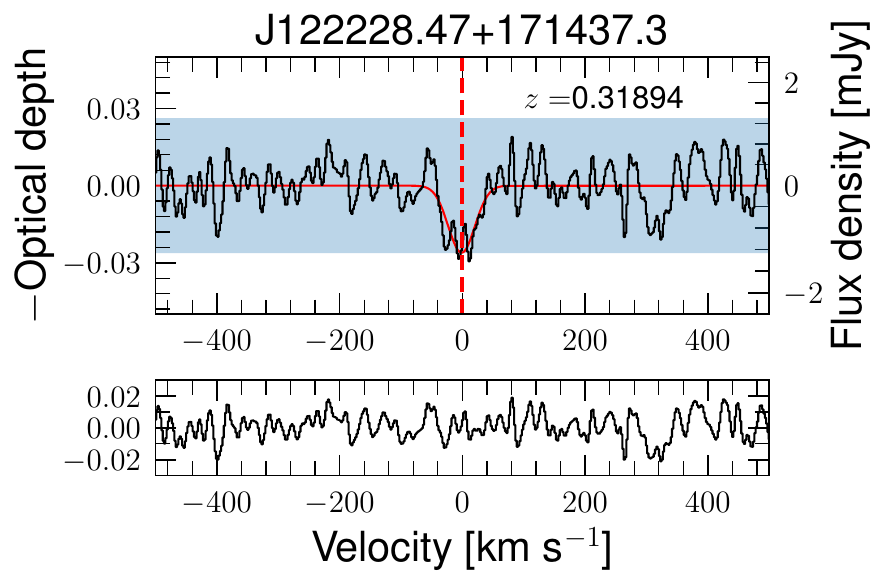}
    \includegraphics[scale=0.50]{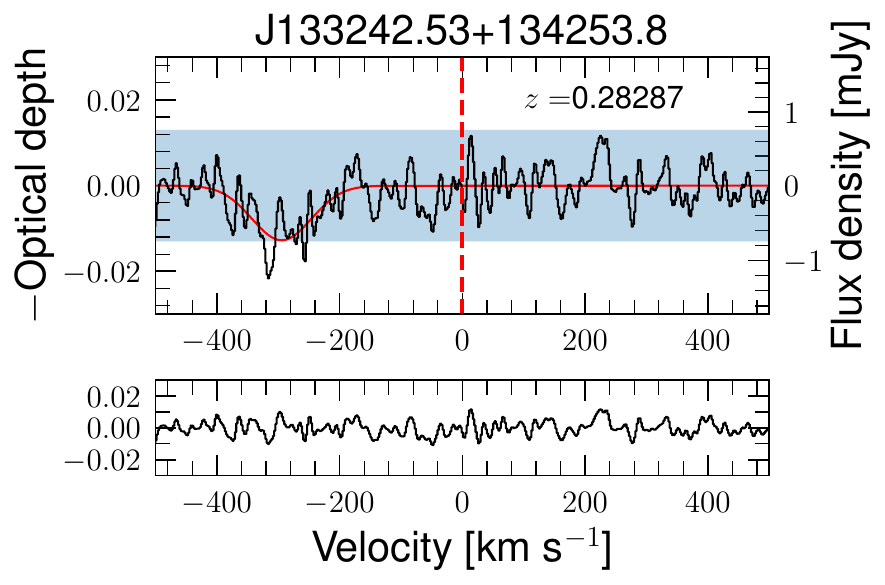}
    \includegraphics[scale=0.50]{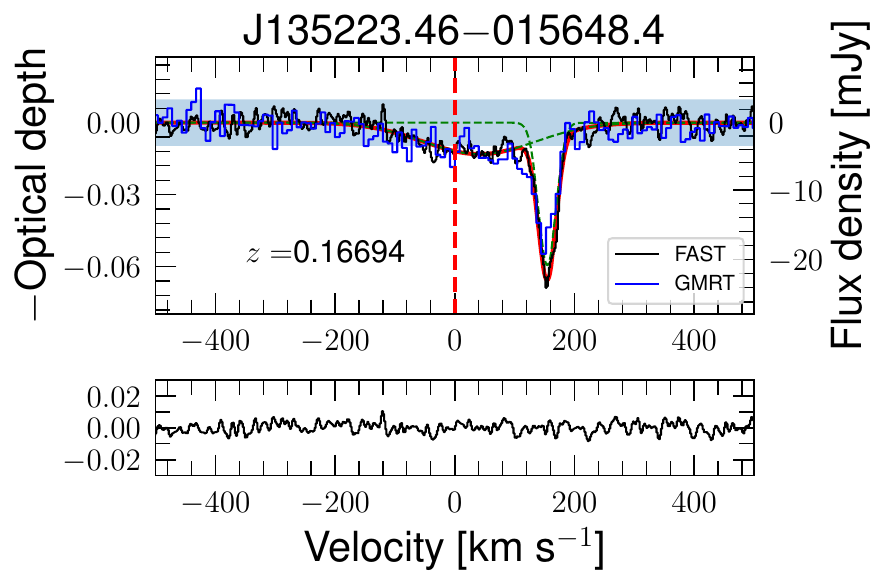}
    \includegraphics[scale=0.50]{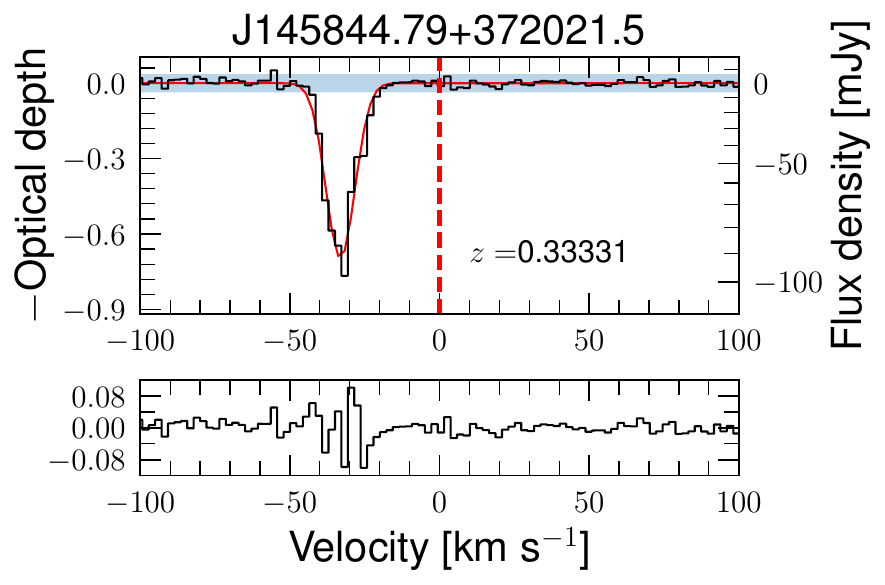}
    \includegraphics[scale=0.50]{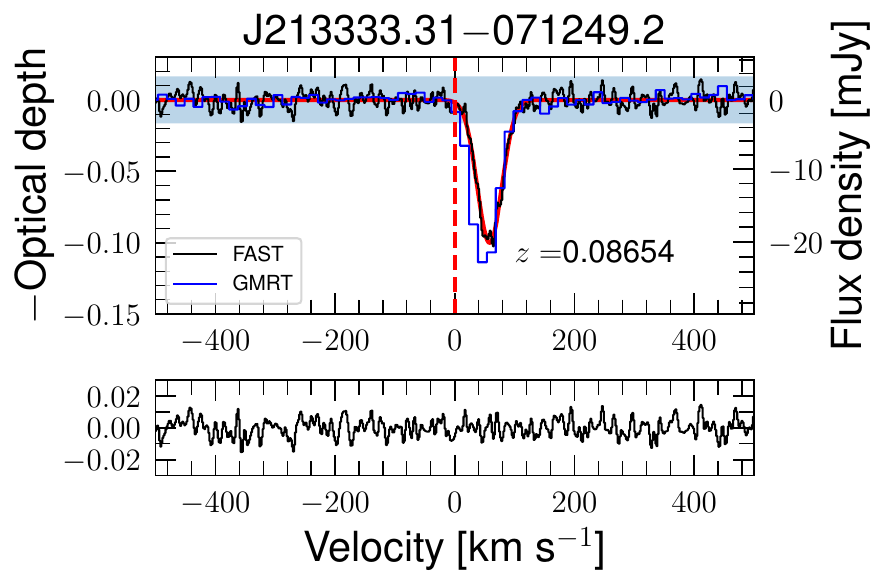}
    \includegraphics[scale=0.50]{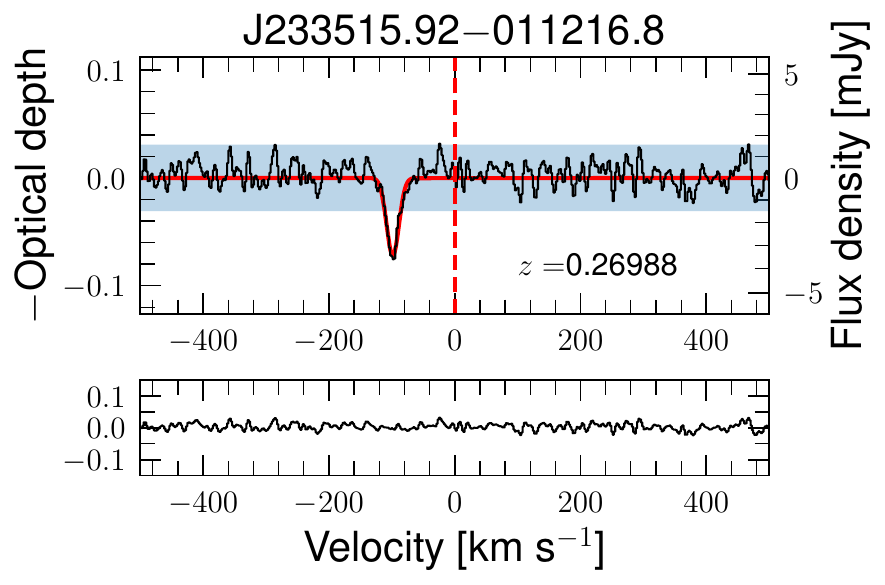}
    \caption{H{\sc i} 21-cm spectra of the sources detected with absorption from FAST observations are shown in black. X-axes represent the velocities w.r.t optical systemic velocity and the Y-axes on the left are optical depths while continuum-subtracted flux densities are shown on the right. The red vertical dashed line at zero marks the optical redshift. The 3$\sigma$ $\tau_{\rm rms}$ values are shown with the blue shades. Grey shades mark the velocities affected by RFI. The Gaussian profile components are shown in green dashed lines while the red solid line shows the combined Gaussian profile. The residuals from the fit are shown at the bottom panel of each plot.}
    \label{fig:hidetections}
\end{figure*}

\begin{figure*}[htb!]
    \centering
    \includegraphics[scale=0.38]{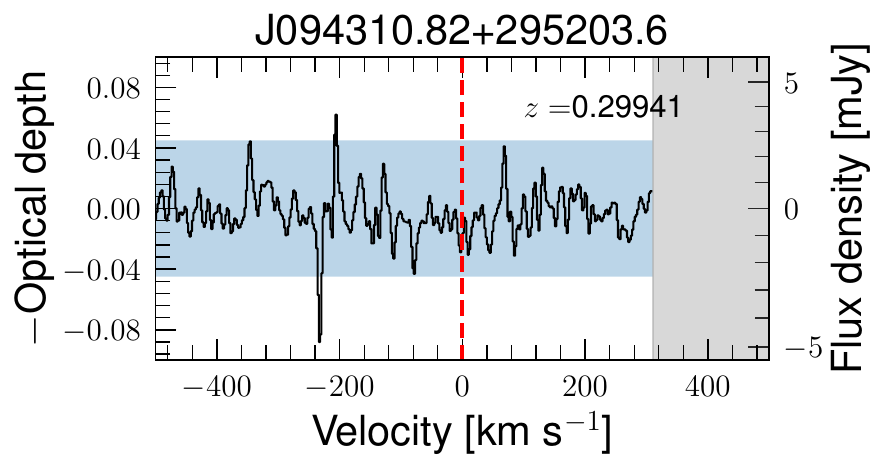}
    \includegraphics[scale=0.38]{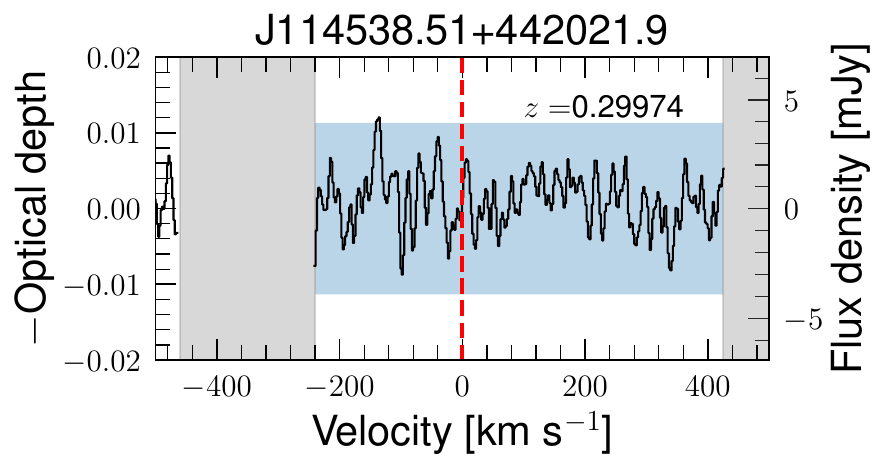}
    \includegraphics[scale=0.38]{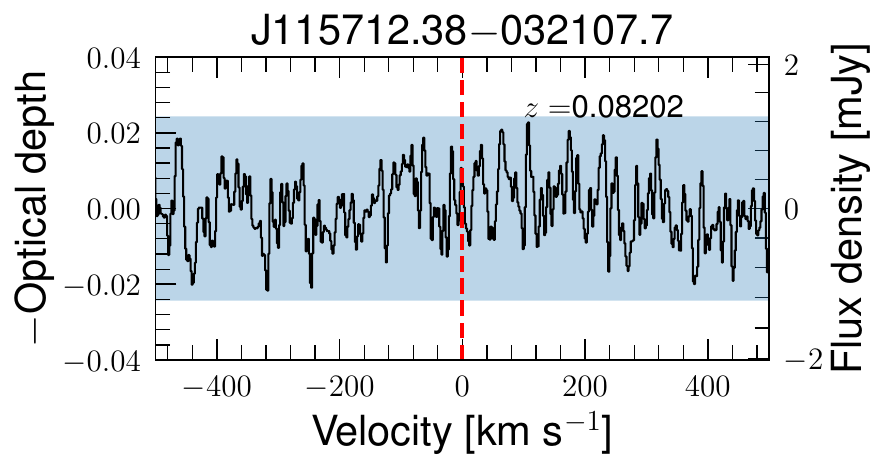}
    \includegraphics[scale=0.38]{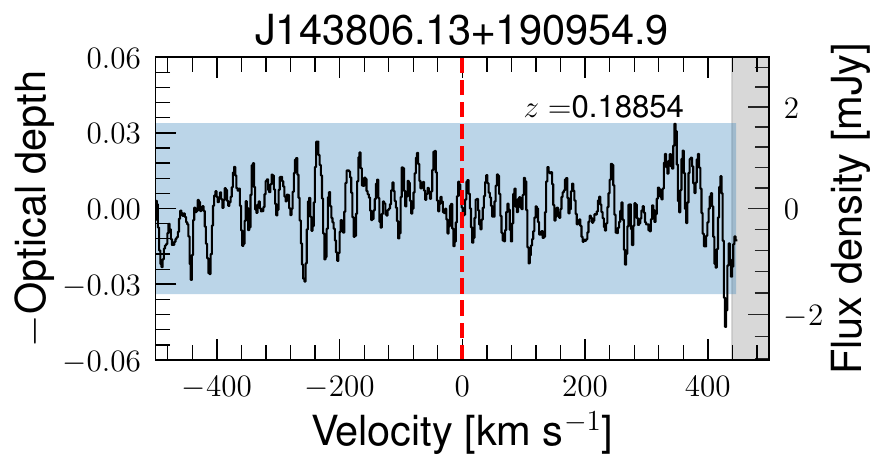}
     \includegraphics[scale=0.38]{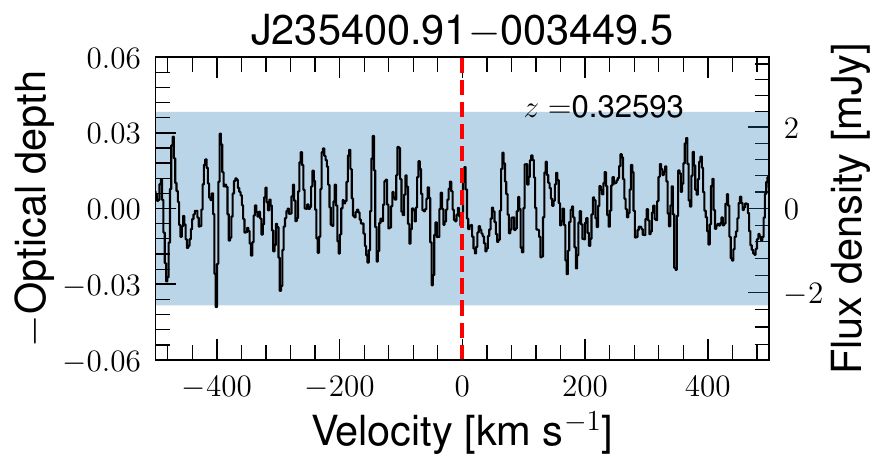}
    \caption{H{\sc i} 21-cm spectra of the sources with no detection of absorption from FAST observations.  X-axes represent the velocities w.r.t optical systemic velocity and the Y-axes on the left are optical depths while continuum-subtracted flux densities are shown on the right. The red vertical dashed line at zero marks the optical redshift. The 3$\sigma$ $\tau_{\rm rms}$ values are shown with the blue shades. Grey shades mark the velocities affected by RFI.}
    \label{fig:hinondetections}
\end{figure*}
\begin{table*}[htb!]
	\caption{Results of the search for associated H{\sc i}  absorption.}
	\begin{center}
		\scriptsize{
			\begin{tabular}{ l c c c c c c c c c c}
				
				\hline
				(1) & (2) & (3) & (4) & (5) & (6) & (7) & (8) & (9) & (10) & (11) \\
				Source name & $S_{\rm c}$(H{\sc i}) &Vel. res. &$\Delta S_{\rm rms}$ (H{\sc i}) & $\tau_{\rm rms}$(H{\sc i}) & $\int{\tau dv}$ & $N$(H{\sc i})& Gaussian  & $V_{\rm shift}$& FWHM & $\tau_{\rm peak}$               \\
                 &mJy bm$^{-1}$& km s$^{-1}$& km s$^{-1}$ & &km s$^{-1}$& 10$^{20}$ cm$^{-2}$& no.  &  km s$^{-1}$& km s$^{-1}$ &   
                    \\
				\hline

SDSS J090410.36+024744.8 & 42.13 &8.2 & 0.58 & 0.0139 & 55.91$\pm$0.98  &101.92$\pm$1.78  & 1  & $-$64.2 &29.1 &0.038 \\
 &  &&  &   &   &  & 2 &$-$36.1  &81.4 &0.401\\
 &  &&  &   &   &  & 3 & $-$1.7 &75.2 &0.115 \\
 &  &&  &   &   & & 4 & 41.2 &51.0 &0.199 \\
SDSS J094310.82+295203.6 & 57.34 &8.4 &0.86  &0.0150  & $<$2.08 &$<$3.79  \\
SDSS J095058.69+375758.8 & 66.56 &6.8 & 0.76 &  0.0114 & 28.50$\pm$0.25  & 51.95$\pm$0.46&1 &54.6 &27.9 & 0.960 \\
SDSS J114538.51+442021.9 & 346.68 &8.3 & 1.31 &0.0038  & $<$0.52 &$<$0.95 & & & &\\
SDSS J115712.38$-$032107.7 & 51.56 & 6.9 &0.42  & 0.0081 & $<$1.02 &  $<$1.86  \\
SDSS J122228.47+171437.3 & 48.93 &11.4 &0.43  & 0.0088&1.41$\pm$0.34 & 2.57$\pm$0.67 & 1 &$-$1.2 &50.9 & 0.026\\
SDSS J133242.53+134253.8 & 57.77 &11.0 &0.25 & 0.0043  & 1.58$\pm$0.25  &2.88$\pm$0.45& 1 &$-$294.4 &114.2 & 0.013  \\
SDSS J135223.46$-$015648.4 &361.2\tablenotemark{\scriptsize{a}} & 7.5 &1.17 & 0.0032 & 5.20$\pm$0.22&9.49$\pm$0.46 & 1 &41.4 &218.7 &0.013 \\
 &  & &  &&& & 2 &154.7 &34.7 &0.059 \\
SDSS J143806.13+190954.9 &48.68 &7.6 & 0.55 & 0.0113  & $<$1.49  &$<$2.72 \\
SDSS J145844.79+372021.5 &183.0\tablenotemark{\scriptsize{b}} &2.1\tablenotemark{\scriptsize{c}} &2.22 &0.0121  & 8.44$\pm$0.04 & 15.39$\pm$0.17 &1 &$-$33.1 &11.4 &0.696 \\
SDSS J213333.31$-$071249.2 &211.3\tablenotemark{\scriptsize{d}} &7.0 & 1.16 & 0.0055 &  5.17$\pm$0.25&9.43$\pm$0.45&1 &58.3 &48.6 & 0.100\\
SDSS J233515.92$-$011216.8 &49.44 &8.1 &0.51 &0.0103 & 1.76$\pm$0.22 & 3.21$\pm$0.41 & 1 & $-$98.5&23.0 &0.072 \\
SDSS J235400.91$-$003449.5 & 61.14 &8.5 &0.78 &0.0128 &$<$1.78& $<$3.25  \\
				\hline
			\end{tabular}
		}
	\end{center}
	\begin{flushleft}
 \tablecomments{
		Column 1: Source name; Column 2: Radio continuum flux density in mJy beam$^{-1}$ at H{\sc i} observing frequency estimated using FIRST peak flux density and spectral index from ~Table\ref{sample}; Column 3: Effective velocity resolution of the spectrum after smoothing;
		Column 4: RMS noise in the absorption spectrum;
		Column 5: 1$\sigma$ optical depth RMS; Column 6: Integrated optical depth or 3$\sigma$ upper limit in km s$^{-1}$;
		Column 7: Column density or 3$\sigma$ upper limit in the units of 10$^{20}$ cm$^{-2}$ with assumptions of spin temperature of $T_{\rm s} =$100 K and covering factor, $f_{\rm c} =$1; Column 8: Gaussian component number; Column 9: Shift in the velocity of Gaussian component relative to optical systemic velocity in km s$^{-1}$; Column 10: full width at half maximum of the Gaussian component in km s$^{-1}$; Column 11: peak optical depth of the Gaussian component.\\
  \tablenotetext{\scriptsize{a}}{We have used the peak flux density at the H{\sc i} observing frequency from our earlier GMRT observations \citep{chandola2020MNRAS.494.5161C} instead of FIRST peak flux density.}
  \tablenotetext{\scriptsize{b}}{Peak flux density value from \cite{murthy2021A&A...654A..94M}.}
  \tablenotetext{\scriptsize{c}}{No smoothing has been done for this profile.}  \tablenotetext{\scriptsize{d}}{Peak flux density value from \cite{chandola2020MNRAS.494.5161C}.}}
	\end{flushleft}
	\label{hiabsresults}
\end{table*}

\begin{figure*}
    \centering
    \includegraphics[scale=0.36]{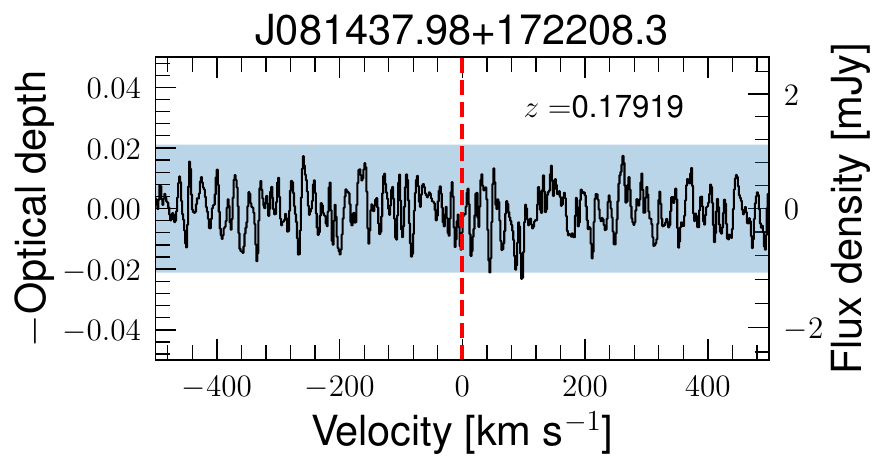} 
    \includegraphics[scale=0.36]{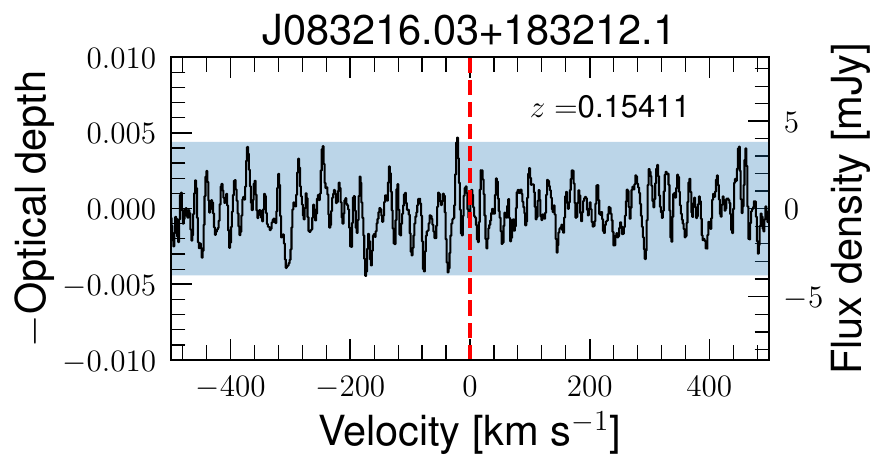} 
    \includegraphics[scale=0.36]{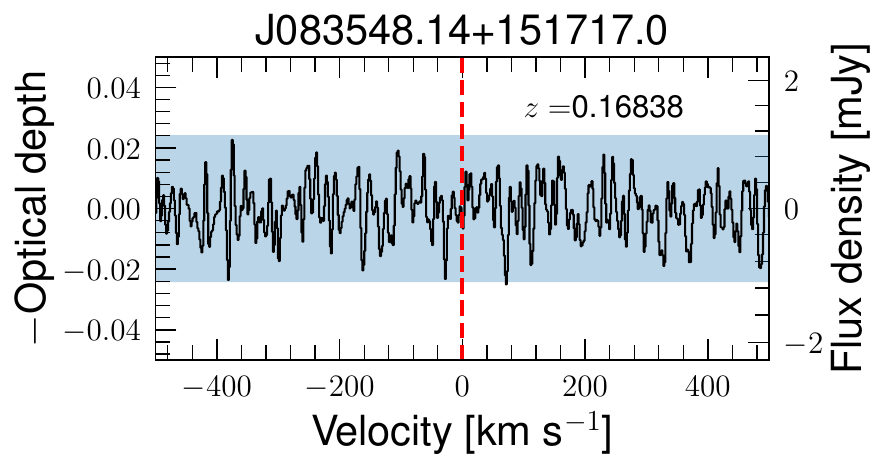}
    \includegraphics[scale=0.36]{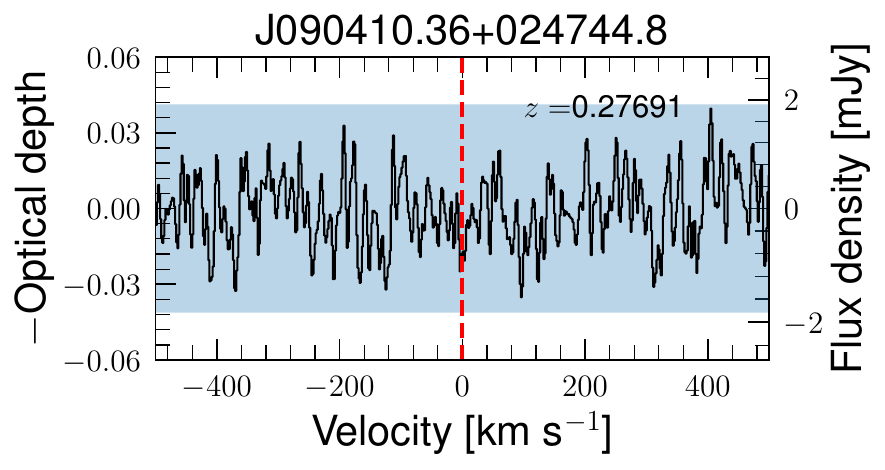}
    \includegraphics[scale=0.36]{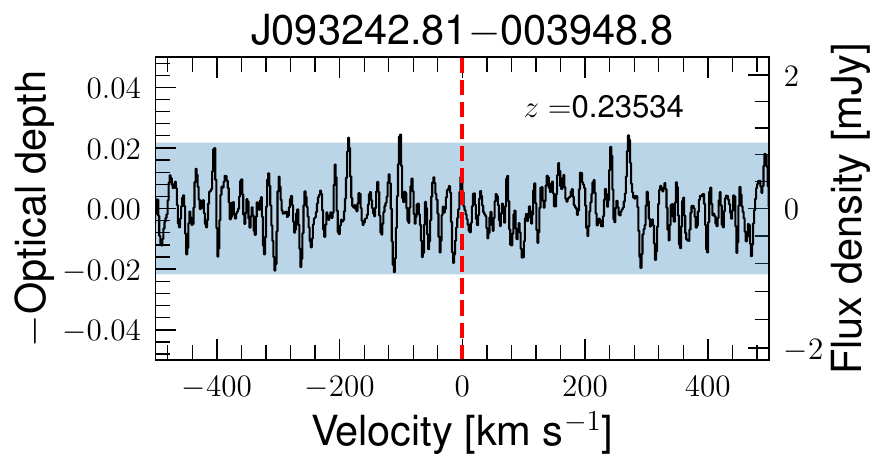}
    \includegraphics[scale=0.36]{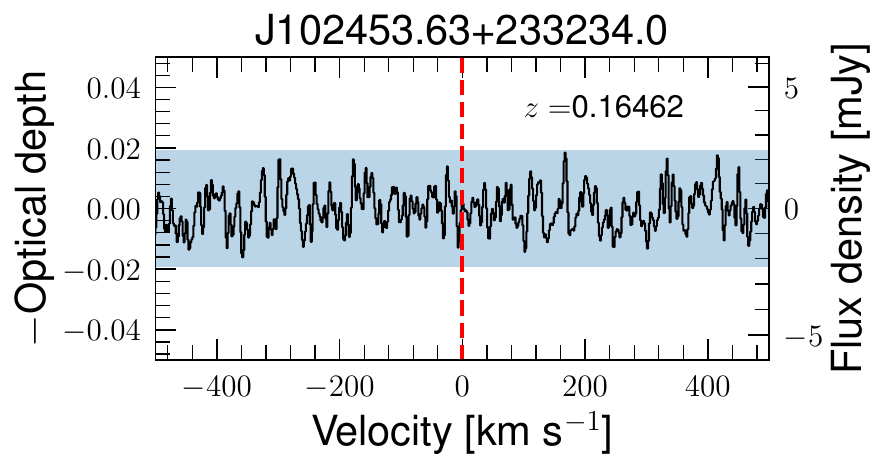}
    \includegraphics[scale=0.36]{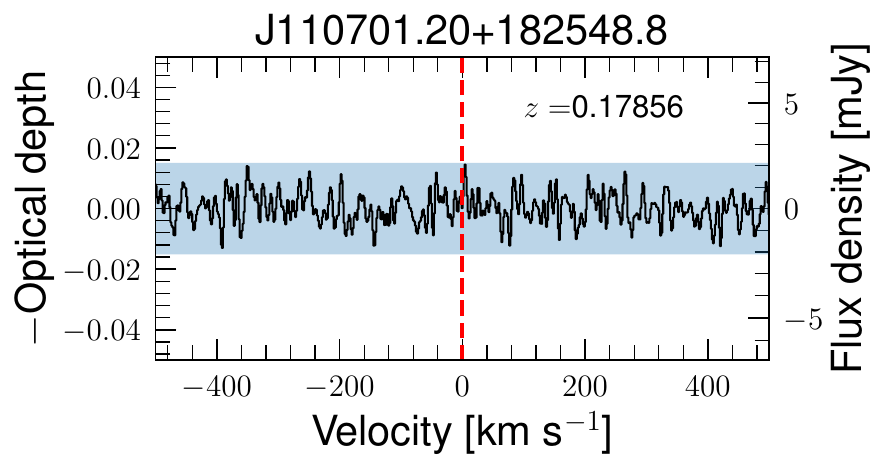}
    \includegraphics[scale=0.36]{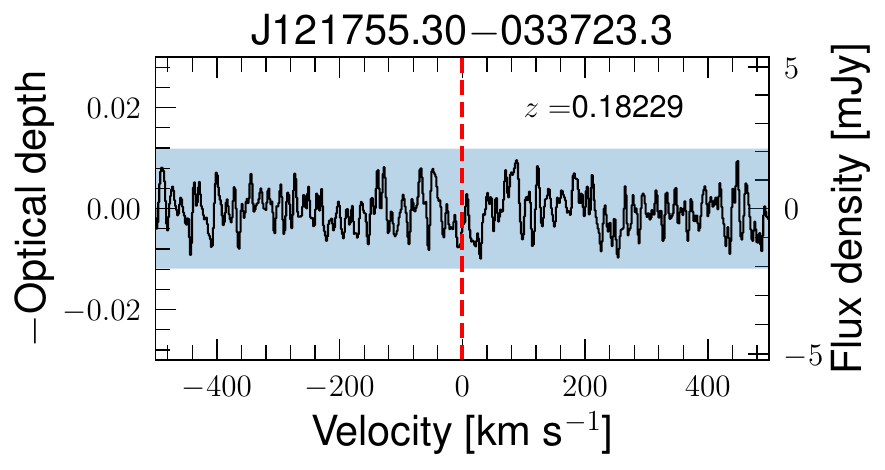}
    \includegraphics[scale=0.36]{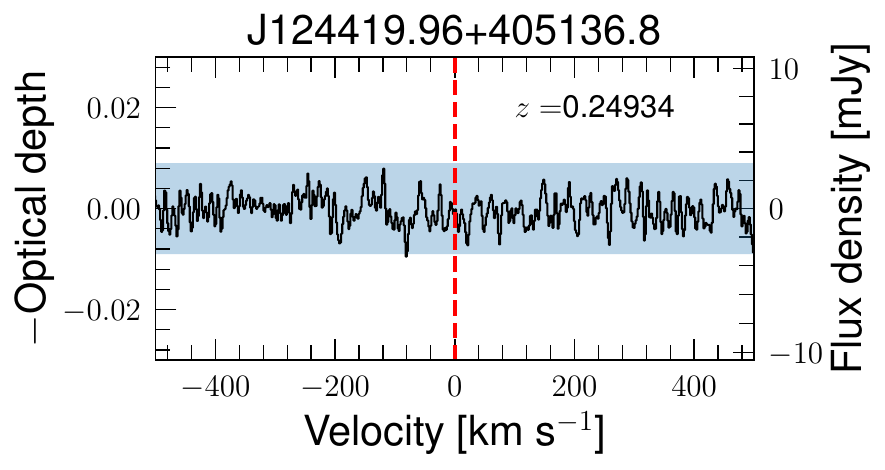}
    \includegraphics[scale=0.36]{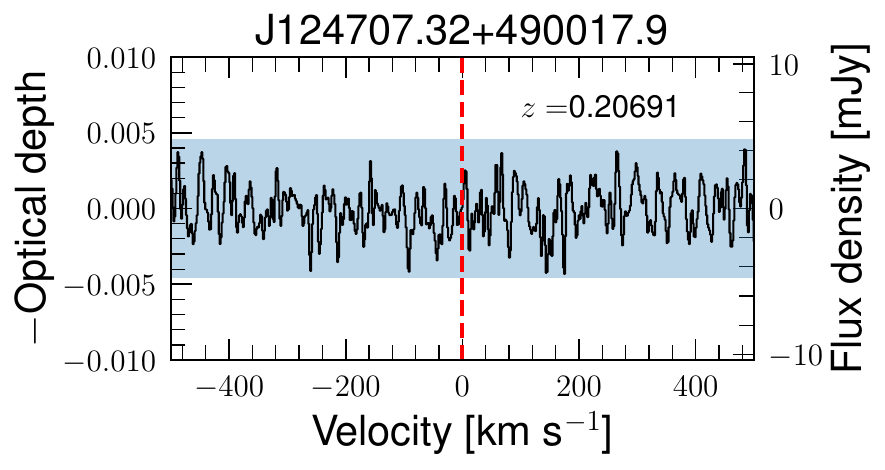}
    \includegraphics[scale=0.36]{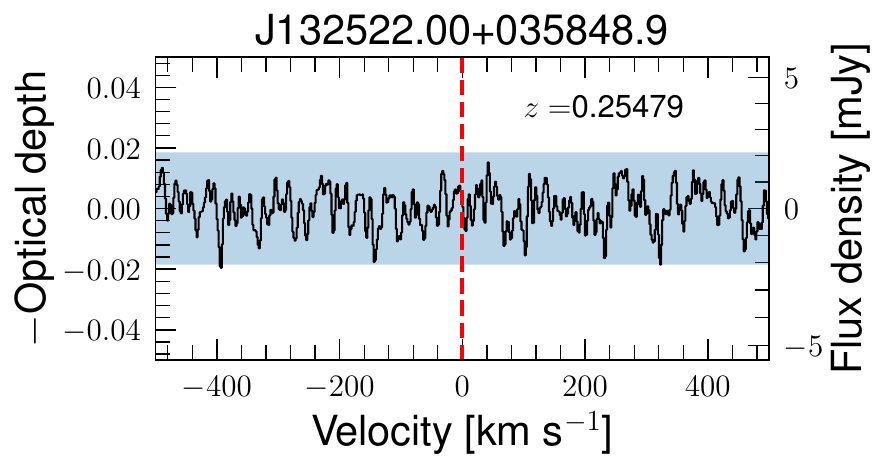}
    \includegraphics[scale=0.36]{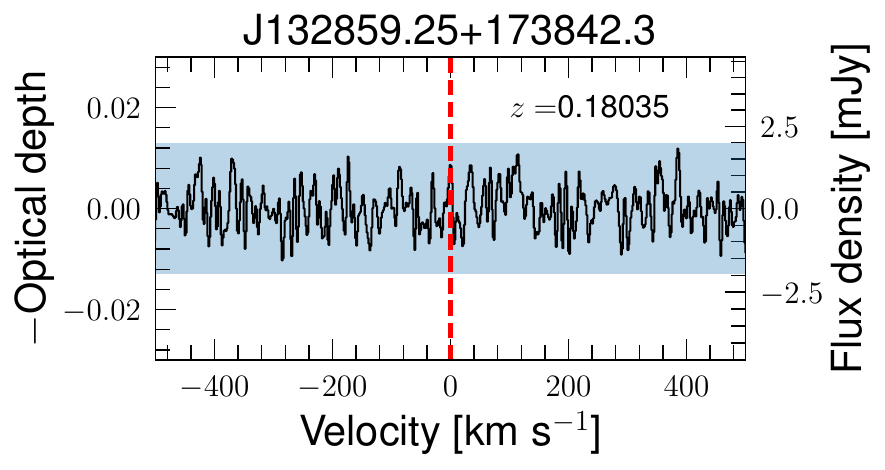}
    \includegraphics[scale=0.36]{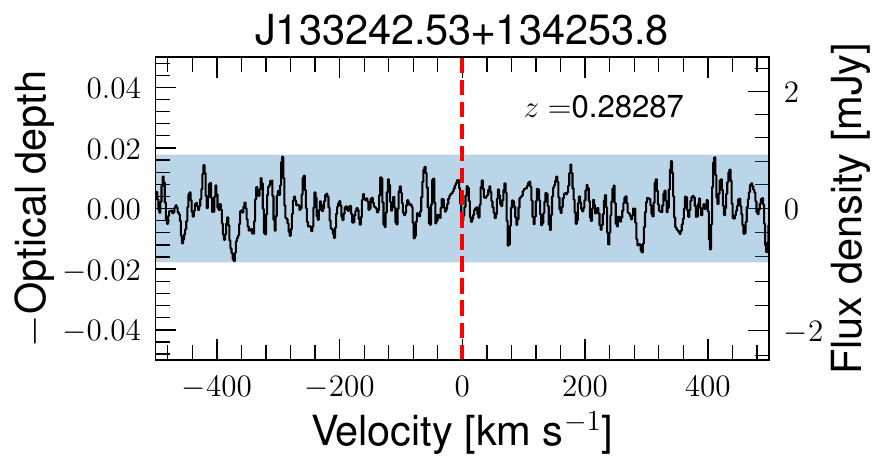}
    \includegraphics[scale=0.36]{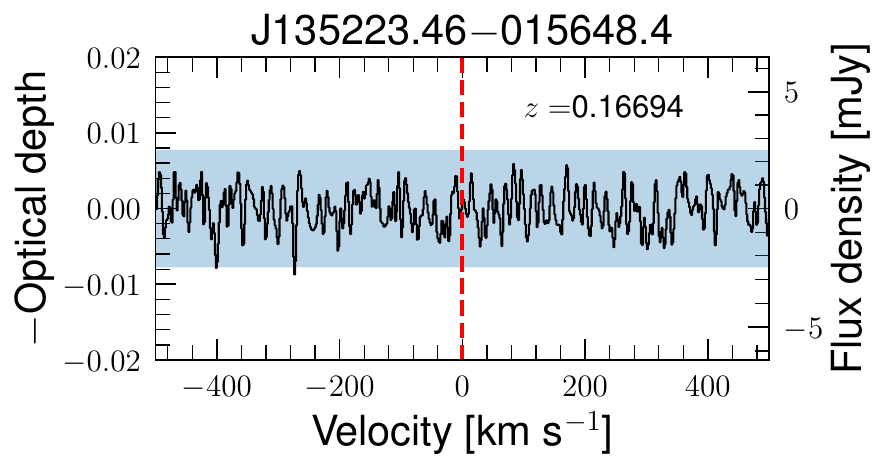}
    \includegraphics[scale=0.36]{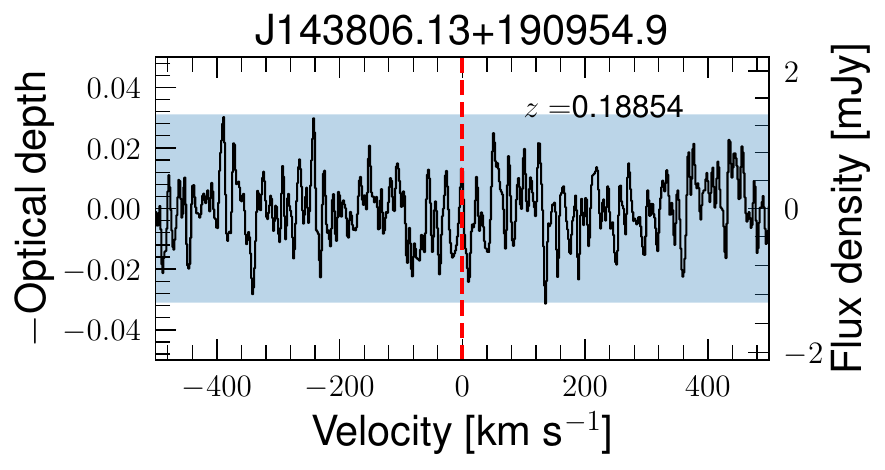}
    \includegraphics[scale=0.36]{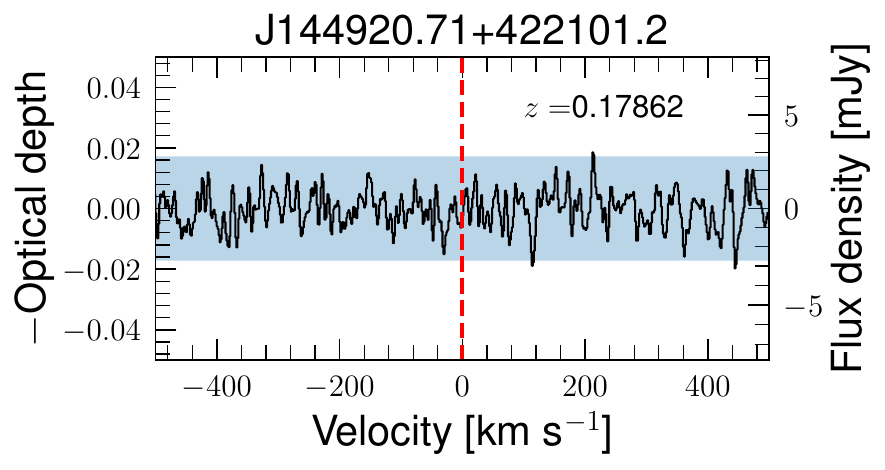}
     \includegraphics[scale=0.36]{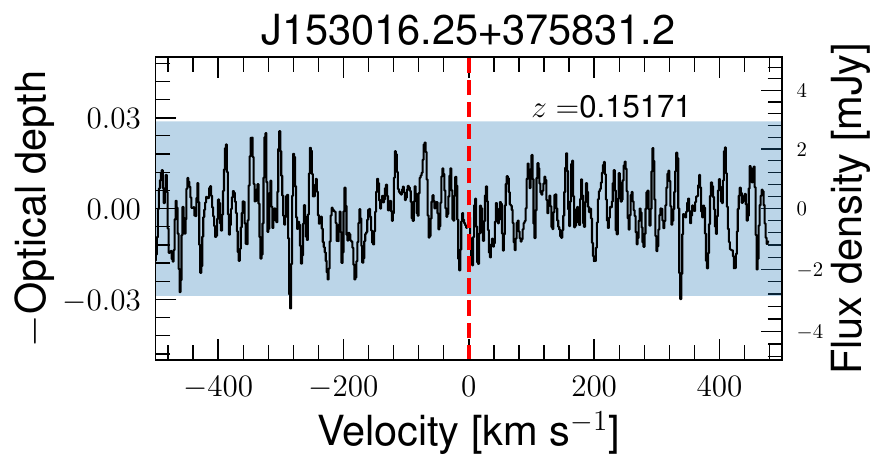}
     \includegraphics[scale=0.36]{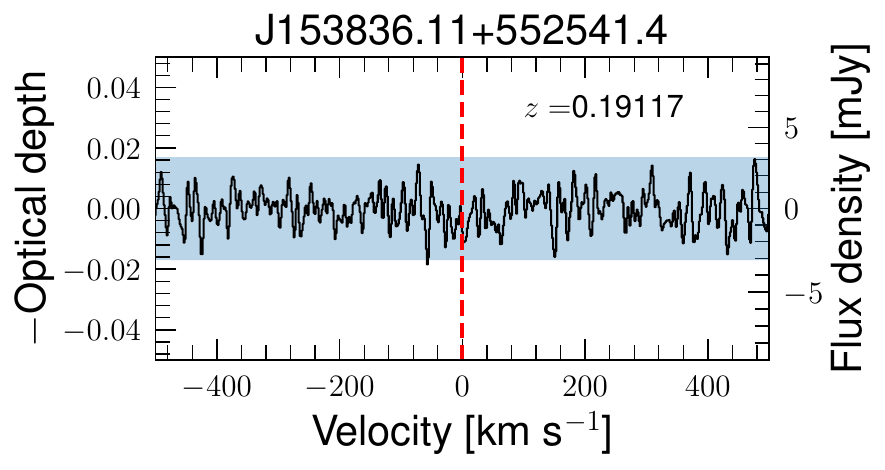}
     \includegraphics[scale=0.36]{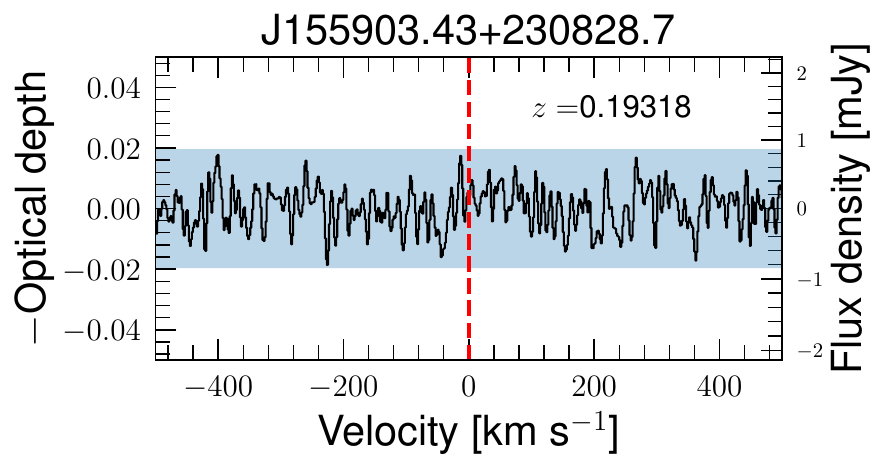}
     \includegraphics[scale=0.36]{J1559+2308OHspecopdhelv.pdf}
      \includegraphics[scale=0.36]{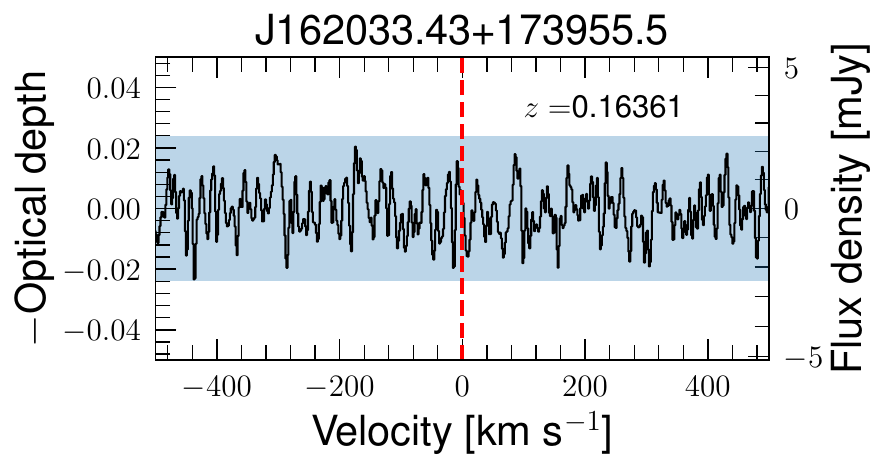}
       \includegraphics[scale=0.36]{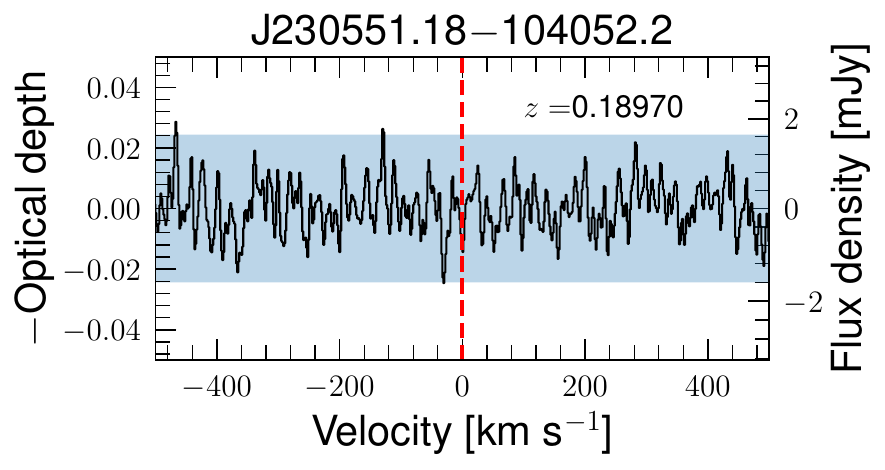}
        \includegraphics[scale=0.36]{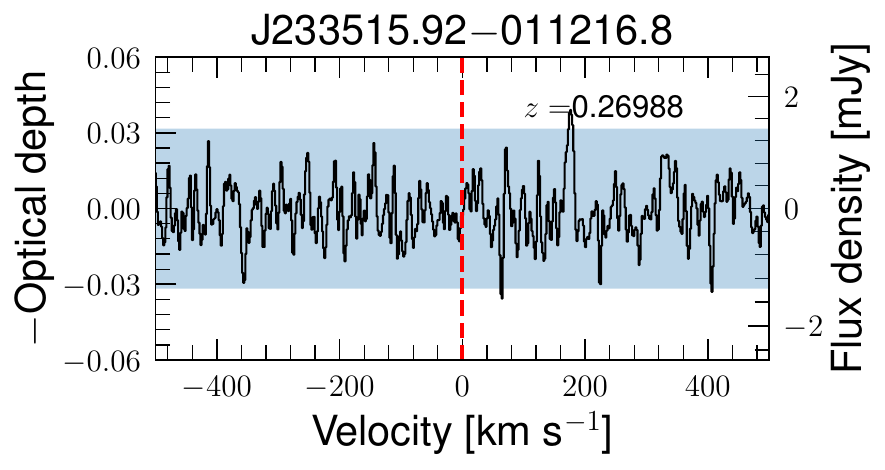}
    \caption{OH 1667 MHz spectra of 23 sources from FAST observations. Symbols mean the same as in Fig.~\ref{fig:hinondetections}.}
    \label{fig:ohnondetections}
\end{figure*}

Of the total 40 sources, we find only 13 sources for which we could either detect H{\sc i} in absorption or set upper limits due to the effects of radio frequency interference (RFI). The rest of the sources were badly affected by RFI from satellites and for one source observations failed due to a problem in the control system. Of these 13 sources, H{\sc i} absorption was detected in 8 sources resulting in a detection rate\footnote{We estimated 1$\sigma$ error on detection rates using \cite{gehrels1986ApJ...303..336G} small number statistics for Poisson distribution.} of 61.5$^{+30.4}_{-21.3}$\%. In Fig.~\ref{fig:hidetections} and Fig.~\ref{fig:hinondetections}, we present the H{\sc i} profiles towards these 13 sources. While three sources, J135223.46$-$015648.4, J145844.79+372021.5 and J213333.31$-$071249.2 were reported earlier in the literature, we find five new cases of H{\sc i} absorption, namely J090410.36+024744.8, J095058.69+375758.8, J122228.47+171437.3, J133242.53+134253.8 and J233515.92$-$011216.8. The detections towards J122228.47+171437.3 and J133242.53+134253.8 are of 3$\sigma$ significance while other detections have $>$ 5$\sigma$ significance. Optical depth from the absorbed flux densities was determined using the equation \citep{morganti2018A&ARv..26....4M},
\begin{equation}
    \tau = -\ln{(1+\frac{\Delta S}{f_c S_{\rm c}})},
\end{equation}
where $\Delta S$, $f_c$ and $S_{\rm c}$ are the continuum subtracted line flux density, covering factor and continuum flux densities at the spectral line frequencies respectively. We assume a covering factor of 1 for our calculations. The median optical depth $\tau_{\rm rms}$(H{\sc i}) is 0.0103 for all 13 sources.
We further fitted the Gaussian function to these profiles to obtain full width at half maximum (FWHM), shift in velocities w.r.t. optical systemic velocities ($V_{\rm shift}$) and peak optical depth ($\tau_{peak}$). We list all the parameters from profiles in Table~\ref{hiabsresults}. We estimated the column densities using the equation \citep{wolfe1975ApJ...200..548W},
\begin{equation}
    N({\rm H\textsc{i}}) = 1.823 \times 10^{18} \hspace{1mm}  T_{\rm s} \hspace{1mm}  \int{\tau dv}  \hspace{1mm} \rm cm^{-2},  
\end{equation}
where we assume spin temperature $T_{\rm s}=$100 K and $\int{\tau dv}$ is the integrated optical depth which for a Gaussian profile is given by 1.064$\times$ FWHM $\times$$\tau_{peak}$. We estimated the errors on integrated optical depth using $\tau_{rms} \times \delta v \times \sqrt{\frac{\rm FWZI}{\delta v}}$. We assume full width zero intensity (FWZI) to be 2.547$\times$FWHM for a Gaussian profile.  The median H{\sc i} column density for H{\sc i} detections has been estimated to be 9.46 $\times$10$^{20}$ cm$^{-2}$. The 3$\sigma$ upper limits on H{\sc i} integrated optical depths were obtained by assuming a Gaussian profile of FWHM$=$100 km s$^{-1}$ and using  3$\times \tau_{rms} \times \delta v \times \sqrt{\frac{\rm FWZI}{\delta v}}$.  The median $N$(H{\sc i}) 3$\sigma$ upper limit for H{\sc i} non-detections has been estimated to be 2.72$\times$10$^{20}$ cm$^{-2}$ which is similar to the median upper limits of 2.6$\times$10$^{20}$ cm$^{-2}$ achieved with our earlier observations  \citep{chandola2020MNRAS.494.5161C} using the Giant Metrewave Radio Telescope (GMRT) or 3.5$\times$10$^{20}$ cm$^{-2}$ by \cite{maccagni2017A&A...604A..43M}.

\begin{table*}[tbh!]
	\caption{Results of the search for associated OH  absorption.}
	\begin{center}
		\scriptsize{
			\begin{tabular}{ l c c c c c c c}
				
				\hline
				(1) & (2) & (3) & (4) & (5) & (6) & (7)& (8) \\
				Source name & $S_{\rm c}$ (OH) &Vel. res. &$\Delta S_{\rm rms}$ (OH) & $\tau_{\rm rms}$(OH) & $\int{\tau dv}$(OH) & $N$(OH) & $N$(OH)/$N$(H{\sc i})    \\
                       &mJy bm$^{-1}$& km s$^{-1}$& km s$^{-1}$ & &km s$^{-1}$& 10$^{15}$ cm$^{-2}$& 10$^{-7}$\\
				\hline

SDSS J081437.98+172208.3 & 51.94&6.5 & 0.37 & 0.0070   &$<$0.86 &$<$1.93&  \\

SDSS J083216.03+183212.1 & 863.11&6.3 & 1.27 & 0.0015   & $<$0.18  &$<$0.40&    \\
SDSS J083548.14+151717.0\tablenotemark{\scriptsize{a}} & 46.31 &6.4 & 0.37  &0.0081 &$<$0.80\tablenotemark{\scriptsize{b}}   & $<$1.79& $<$12.2     \\
SDSS J090410.36+024744.8\tablenotemark{\scriptsize{a}} & 45.43 &7.0 & 0.63  &0.0137   &$<$1.70   & $<$3.80& $<$3.7  \\

SDSS J093242.81$-$003948.8 &44.48 &6.8 & 0.32 & 0.0072 & $<$0.90 & $<$2.03&  \\
SDSS J102453.63+233234.0 &122.04 &6.4 & 0.78  & 0.0064  &$<$0.78 &$<$1.74&  \\

SDSS J110701.20+182548.8 &140.89 &6.5 & 0.71  & 0.0050   &$<$0.61 & $<$1.38&  \\
SDSS J121755.30$-$033723.3 &175.87 &6.5 & 0.70  & 0.0040 & $<$0.48& $<$1.08&  \\
SDSS J124419.96+405136.8 &355.79 &6.9 & 1.07  & 0.0030  & $<$0.38 & $<$0.84& \\
SDSS J124707.32+490017.9\tablenotemark{\scriptsize{a}} &1043.26 &6.6 & 1.59  & 0.0015 & $<$0.28\tablenotemark{\scriptsize{b}} & $<$0.62&$<$73.9 \\
SDSS J132522.00+035848.9 &113.11 &6.9 & 0.70 & 0.0062  & $<$0.78 & $<$1.74&\\
SDSS J132859.25+173842.3 &152.36 &6.5 & 0.66 & 0.0043 & $<$0.53  & $<$1.18&  \\
SDSS J133242.53+134253.8\tablenotemark{\scriptsize{a}} &50.70 &7.0 &0.30 &  0.0059 & $<$0.80 &$<$1.80&$<$62.5 \\
SDSS J135223.46$-$015648.4\tablenotemark{\scriptsize{a}} &321.88\tablenotemark{\scriptsize{c}} &6.4 &0.83 & 0.0026 & $<$0.41 & $<$0.92&$<$9.7\\

SDSS J143806.13+190954.9 &43.09 &6.5 & 0.45 & 0.0104  & $<$1.26  & $<$2.83&\\
SDSS J144920.71+422101.2 &159.71 &6.5 & 0.92 & 0.0057  & $<$0.70 & $<$1.57& \\
SDSS J153016.25+375831.2 &100.60 &6.3 & 0.96 & 0.0096 & $<$1.15 & $<$2.58& \\
SDSS J153836.11+552541.4\tablenotemark{\scriptsize{a}} &184.24 &6.5  & 1.04 & 0.0057  & $<$0.77\tablenotemark{\scriptsize{d}} & $<$1.73&$<$48.1 \\
SDSS J155903.43+230828.7 &43.69 &6.5  & 0.29 & 0.0066 & $<$0.80 & $<$1.79& \\
SDSS J155927.67+533054.4 &168.80 &6.5 & 1.20 & 0.0071  & $<$0.86 & $<$1.94&  \\
SDSS J162033.43+173955.5 &104.86 &6.4 & 0.83 & 0.0080 &$<$0.96  &$<$2.16& \\
SDSS J230551.18$-$104052.2 & 66.41 &6.5 & 0.54 & 0.0081 & $<$0.99 & $<$2.22& \\
SDSS J233515.92$-$011216.8\tablenotemark{\scriptsize{a}} & 44.06&7.0 &0.47 & 0.0106& $<$0.64 & $<$1.44 & $<$44.9 \\
				\hline
			\end{tabular}
		}
	\end{center}
\begin{flushleft}
\tablecomments{Column 1: source name; Column 2: Radio continuum flux density in mJy beam$^{-1}$ at OH observing frequency estimated using FIRST peak flux density and spectral index from ~Table\ref{sample} except for J083548.14+151717.0 where we don't have the spectral index and use the FIRST peak flux density;  Column 3: Effective velocity resolution of the spectrum after smoothing;
		Column 4: RMS noise in the absorption spectrum;
		Column 5: 1$\sigma$ optical depth RMS; Column 6: 3$\sigma$ upper limit on integrated optical depth assuming line full-width half maximum (FWHM) of 100 km s$^{-1}$ except for the sources with associated H{\sc i} absorption where we use the full width at zero intensity (FWZI) of the full H{\sc i} absorption profiles;
		Column 7:  3$\sigma$ upper limit on OH column densities in the units of 10$^{15}$ cm$^{-2}$ with assumption of excitation temperature of $T_{\rm ex} =$10 K and covering factor, $f_{\rm c}$(OH) $=$1; Column 8: Upper limits on $N$(OH)/$N$(H{\sc i}) in the units of 10$^{-7}$ for sources with associated H{\sc i} absorption.\\
  \tablenotetext{\scriptsize{a}}{Sources detected with associated H{\sc i} absorption.}
  \tablenotetext{\scriptsize{b}}{Used $\int\tau dv$ (H{\sc i}) and $\tau_{\rm peak}$(H{\sc i}) from \cite{maccagni2017A&A...604A..43M} to obtain FWHM(H{\sc i}) for a single Gaussian profile and converted to FWZI using FWZI$=$2.547$\times$FWHM.} 
  \tablenotetext{\scriptsize{c}}{Continuum flux density extrapolated to OH frequency using the flux density value for H{\sc i} frequency from Table~\ref{hiabsresults} and spectral index from Table~\ref{sample}.}}
  \tablenotetext{\scriptsize{d}}{Using  Gaussian profile parameters from \citet{chandola2020MNRAS.494.5161C} to estimate FWZI.}
	\end{flushleft}
	\label{ohabsresults}
\end{table*}

In addition to H{\sc i} absorption, we have 23 spectra unaffected with the RFI at their OH observing frequencies (shown in Fig.~\ref{fig:ohnondetections}). We obtained a median $\tau_{\rm rms}$ (OH) of 0.0064. We did not detect OH absorption in any of these sources, although it is possible in the source J133242.53+134253.3 at a blue-shifted velocity of $\sim$400 km s$^{-1}$ which needs to be probed from deeper observations. We estimated the upper limits on the OH column densities using the equation \citep{liszt1996A&A...314..917L},
\begin{equation}
    N({\rm OH}) = 2.24 \times 10^{14} \hspace{1mm}  T_{\rm ex} \hspace{1mm}  \int{\tau dv}  \hspace{1mm} \rm cm^{-2},  
\end{equation}
where we assume excitation temperature ($T_{\rm ex}$)$=$ 10 K and obtain integrated optical depth using FWZI assuming line FWHM$=$100 km s$^{-1}$.  For the sources detected with H{\sc i} absorption, we use the FWZI of the full H{\sc i} absorption lines. For the sources with multiple Gaussian components, we determined the FWZI of the full profile using ($V_{\rm high}$+FWZI$_{\rm high}$/2)$-$($V_{\rm low}$$-$ FWZI$_{\rm low}$/2), where \lq high' and \lq low' are the Gaussian components at the highest and lowest velocities.  For all OH non-detections, we achieved a median 3$\sigma$ upper limit of 1.74$\times$10$^{15}$ cm$^{-2}$. This median upper limit on OH column density is similar to the median OH column density value of 1.92$\times$10$^{15}$ cm$^{-2}$ for 4 intervening OH absorbers reported in the literature \citep{kanekar2002A&A...381L..73K,kanekar2003A&A...412L..29K,kanekar2012ApJ...746L..16K,gupta2018ApJ...860L..22G}. We have listed the upper limits for all 23 sources in Table~\ref{ohabsresults}.

\begin{figure}[htb!]
    \centering
    \includegraphics[scale=0.5]{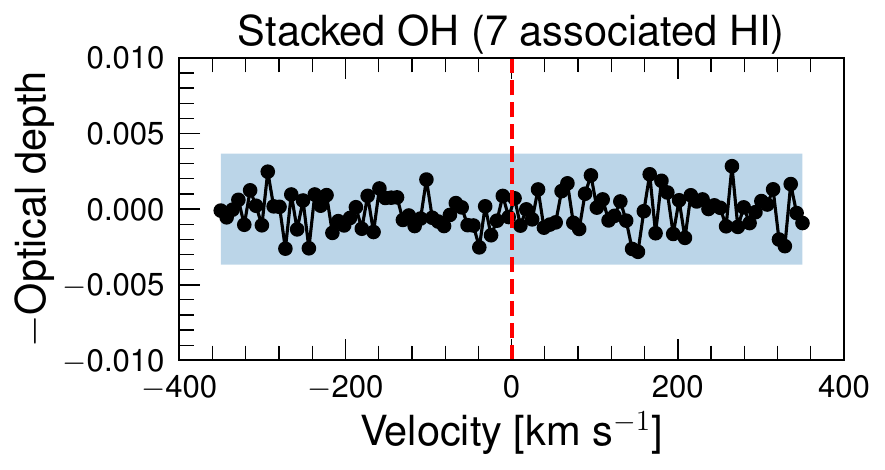}
    \includegraphics[scale=0.5]{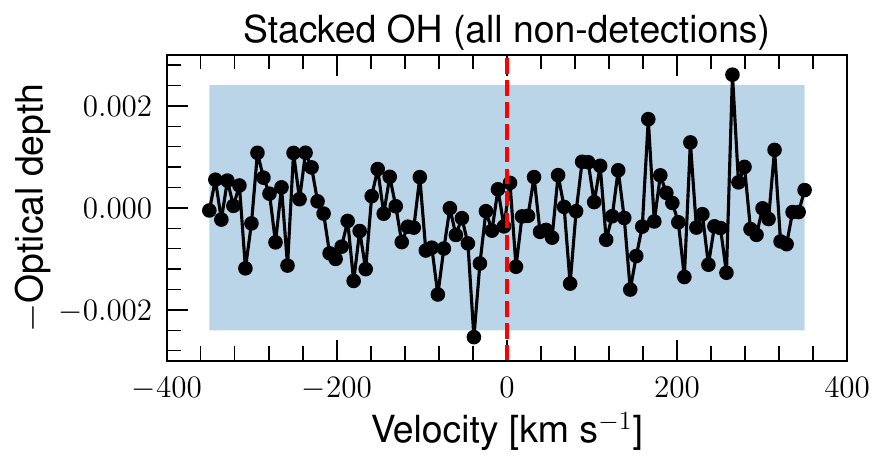}
    \caption{Top panel: Stacked OH spectrum for seven known associated H{\sc i} absorbers, 083548.14+151717.0, J090410.36+024744.8, J124707.32+490017.9, J135223.46$-$015648.4, J133242.53+134253.8, J153836.11+552541.4 and J233515.92$-$011216.8. We obtained $\tau_{\rm rms}$(OH) value of 0.00122 for this spectrum for velocity resolution of 7 km s$^{-1}$ corresponding to the 3$\sigma$ upper limit on $N$(OH) of 3.47$\times$10$^{14}$ cm$^{-2}$. Bottom panel: Stacked OH spectrum for all the non-detections shown in Fig.~\ref{fig:ohnondetections}. The $\tau_{\rm rms}$(OH) value for this spectrum is 0.0008 which corresponds to the 3$\sigma$ upper limit on $N$(OH) of 2.27$\times$10$^{14}$ cm$^{-2}$.}
    \label{OHstackedhi}
\end{figure}

 Of these 23 sources, we have H{\sc i} absorption data for  5 sources, J090410.36+024744.8, J133242.53+134253.8, J135223.46$-$015648.4, J143806.13+190954.9  and J233515.92$-$011216.8. These include the detections, J090410.36+024744.8, J133242.53+134253.8, J135223.46$-$015648.4 and J233515.92$-$011216.8.  We achieved a median 3$\sigma $ upper limit of 1.62 $\times$10$^{15}$ cm$^{-2}$ on OH column density of these four sources. For the source J153836.11+552541.4, although due to the presence of RFI, we could not confirm H{\sc i} absorption we had earlier detected with the GMRT \citep{chandola2020MNRAS.494.5161C}, we have RFI-free OH spectra in this paper. Similarly, for the sources J083548.14+151717.0 and J124707.32+490017.9, \cite{maccagni2017A&A...604A..43M} had reported H{\sc i} absorption which we could not detect due to RFI but have OH spectra.
We further stacked the OH absorption spectra of these 7 associated H{\sc i} absorbers detections and all OH non-detections separately (see Fig.~\ref{OHstackedhi}) by normalising using the variance i.e. 1/$\tau_{\rm rms}^{2}$, in these profiles. For the stacked OH spectrum of seven H{\sc i} absorbers, we obtained $\tau_{\rm rms}$(OH) of 0.00122, which corresponds to a 3$\sigma$ upper limit on OH column density of 3.47$\times$10$^{14}$ cm$^{-2}$. Using the mean H{\sc i} column density, 19.5$\times$10$^{20}$ cm$^{-2}$, for seven H{\sc i} absorbers, we estimated an [OH]/[H{\sc i}] upper limit of 1.78$\times$10$^{-7}$. For the stacked OH spectra of all OH non-detections, we obtained  $\tau_{\rm rms}$(OH) of 0.0008, which corresponds to a 3$\sigma$ upper limit on OH column density of 2.27$\times$10$^{14}$ cm$^{-2}$.

\subsection{Individual sources with H{\sc i} absorption}
\subsubsection{New sources with H{\sc i} absorption detections}
\emph{J090410.36+024744.8:} 
We detected H{\sc i} absorption towards this source at a redshift of 0.27691. This source has been classified as a LERG by \cite{best2012MNRAS.421.1569B}. The peak flux density in FIRST is 46.94 mJy beam$^{-1}$ and the integrated flux density is 48.26 mJy. This source is unresolved in FIRST. It has a radio luminosity of $\sim$10$^{25}$ W Hz$^{-1}$ at 1.4 GHz. The radio continuum spectrum shows inversion ($\alpha_{\rm 150 MHz}^{\rm 1.4 GHz}<$ 0) at lower frequencies as it is not detected in the TGSS image. The H{\sc i} absorption profile has four Gaussian components, three blueshifted and one redshifted, all within $\pm$100 kms$^{-1}$ of optical redshift. The multiple-component profile suggests the complex motion of the gas. The H{\sc i} column density has been estimated to be (101.92$\pm$1.78)$\times$10$^{20}$ cm$^{-2}$.

\emph{J095058.69+375758.8:} J095058.69+375758.8 is a Seyfert 1 galaxy at a redshift of 0.04053 and classified as HERG by \cite{best2012MNRAS.421.1569B}. The optical SDSS image shows an edge-on disk-like morphology. The radio power of the source is $\sim$10$^{23.4}$ W Hz$^{-1}$ with a peak flux density of 67.5 mJy beam$^{-1}$ in FIRST. The continuum radio spectrum shows inversion at lower frequencies with no detection in the TGSS image. We detect H{\sc i} emission as well as absorption towards this source. The H{\sc i} absorption component is a narrow (FWHM$\sim$29.9 km s$^{-1}$) and deep ($\tau_{peak} \sim$0.9) profile redshifted by $\sim$54.6 kms$^{-1}$ relative to optical systemic velocity. The H{\sc i} column density for the H{\sc i} absorbing component  has been estimated to be (51.90$\pm$0.46) $\times$10$^{20}$ cm$^{-2}$.

\emph{J122228.47+171437.3:} 
This radio source is a HERG at a redshift of 0.31894 corresponding to H{\sc i} observing frequency of 1076.9298 MHz. It has a flat continuum radio spectrum ($\alpha_{\rm 150 MHz}^{\rm 1.4 GHz} \sim$0.14). This source shows an absorption feature in the profile at optical redshift though at 3$\sigma$.  We estimate an H{\sc i} column density of (2.57$\pm$0.67) $\times$10$^{20}$ cm$^{-2}$.
 
 \emph{ J133242.53+134253.8:} This source is a HERG at a redshift of 0.28287 corresponding to the expected H{\sc i} frequency of 1107.2094 MHz. It has a steep spectrum with a spectral index $\alpha_{\rm 150 MHz}^{\rm 1.4 GHz} \sim$0.82. It could be a compact steep spectrum or extended radio source not resolved in the FIRST image. The SDSS optical image shows a possible tidal tail-like merger feature. The H{\sc i} spectrum shows a shallow absorption feature (optical depth $\sim$0.013) of full-width half maximum of 114.2 km s$^{-1}$ at around $-$294 kms$^{-1}$ relative to optical systemic velocity, though at 3$\sigma$ significance. The H{\sc i} column density has been estimated to be (2.88$\pm$0.45)$\times$10$^{20}$ cm$^{-2}$. \\
 %

\emph{J233515.92$-$011216.8 :}
We detected H{\sc i} absorption towards this radio source at a redshift of 0.26988. This source is classified as a LERG by \cite{best2012MNRAS.421.1569B} and has a radio luminosity of $\sim$10$^{24.8}$W Hz$^{-1}$ with a steep low-frequency spectrum ($\alpha_{\rm 150 MHz}^{\rm 1.4 GHz} \sim$0.72). The H{\sc i} absorption is a narrow (FWHM$\sim$21.8 km s$^{-1}$) profile blue-shifted by $\sim$99 km s$^{-1}$ relative to the optical systemic velocity. The H{\sc i} column density has been estimated to be (3.21$\pm$0.41)$\times$10$^{20}$cm$^{-1}$.

\subsubsection{H{\sc i} absorption detection towards radio sources previously reported with H{\sc i} absorption}
\emph{J135223.46$-$015648.4:} We had reported H{\sc i} absorption towards this source, a HERG at the redshift of 0.167, in our earlier paper \citep{chandola2020MNRAS.494.5161C}. This source has an H{\sc i} absorption profile with two components one broad and shallow, and another deep and narrow component. To make a comparison of the profiles obtained with the GMRT and FAST, we plot the profiles together in Fig.~\ref{fig:hidetections}. This plot shows that the profiles are consistent from two different observations and FAST can detect shallow profiles of optical depth $\sim$0.006 in a relatively short time. 

This source has an extended resolved radio structure in the observations with GMRT. Since the profiles from different spatial resolutions are similar, we conclude that the H{\sc i} absorption is from the compact radio component. We used the peak flux density from GMRT observations to estimate the optical depth. We estimate an H{\sc i} column density of (9.49$\pm$0.46)$\times$10$^{20}$ cm$^{-2}$ from FAST observations which is consistent within 1$\sigma$ with our earlier GMRT results. 

\emph{J145844.79+372021.5:} This source is classified as a LERG by \citep{best2012MNRAS.421.1569B} at the redshift of 0.33331$\pm$0.00008. The low-frequency radio continuum spectrum of the source shows an inversion ($\alpha_{\rm 150 MHz}^{\rm 1.4 GHz} \sim -$0.31) indicating a possibility of a  Peaked spectrum (PS) source \citep{odea2021A&ARv..29....3O}. However, this source was earlier classified as a flat-spectrum radio source based on a spectral index of $-$0.46 between 1.4 GHz and 4.85 GHz by \citet{taylor1996ApJS..107...37T}. The very long baseline interferometry (VLBI) scale maps show a very compact radio structure of $\lesssim$10 milliarcseconds corresponding to a projected linear size of $\lesssim$ 48 parsecs \citep{helmboldt2007ApJ...658..203H,murthy2021A&A...654A..94M}. This source is classified as a blazar and could be a BL Lac object \citep{dbrusco2014ApJS..215...14D}.

This source had been reported with H{\sc i} absorption earlier by \cite{aditya2018MNRAS.481.1578A} and \cite{murthy2021A&A...654A..94M}. The profile has a single deep (optical depth $\sim$0.70) and narrow component (FWHM $\sim$11.4 km s$^{-1}$) reflecting an absorption due to the H{\sc i} disk. Although \cite{aditya2018MNRAS.481.1578A} and \cite{murthy2021A&A...654A..94M} found this profile to be blueshifted by $\sim$60 km s $^{-1}$ and $\sim$40 km s$^{-1}$ relative to the optical redshift, the profile from FAST observations show a smaller shift relative to optical systemic velocity with a blueshift of 33.1 km s$^{-1}$. It is to be noted that this difference could be due to the different redshift of 0.33343 \citep{schneider2005AJ....130..367S} used by \cite{aditya2018MNRAS.481.1578A}. We also notice that the profile is deeper relative to earlier observations by \cite{aditya2018MNRAS.481.1578A} and \cite{murthy2021A&A...654A..94M} resulting in higher column density value $\sim$15.4 $\times$10$^{20}$cm$^{-2}$. 
This difference in the absorption profile depth could result from the intrinsic variability of the source, with the compact radio component intersecting clouds of different depths during the different observation epochs. 

\emph{ J213333.31$-$071249.2:} This source is a HERG at a redshift of 0.087 and had been reported with H{\sc i} absorption in an earlier paper by us \citep{chandola2020MNRAS.494.5161C}. It has a steep low-frequency radio continuum spectrum ($\alpha_{\rm 150 MHz}^{\rm 1.4 GHz} \sim$0.57).  We compare the H{\sc i} absorption profiles from GMRT and FAST. The two profiles are consistent with a slight offset in velocity but less than the spectral resolution. Although the GMRT profile appears slightly deeper, it is within the 10\% error on gain calibration. We estimate the column density value to be (9.43$\pm$0.45)$\times$10$^{20}$ cm$^{-2}$.

\subsubsection{H{\sc i} absorption non-detections}

 \emph{J094310.82+295203.6:} This radio  AGN is identified as a LERG by \cite{best2012MNRAS.421.1569B} at the SDSS optical redshift of 0.29941. The H{\sc i} spectrum is partially affected due to RFI at 1090 MHz, requiring further checking. However, we categorise this source as a non-detection with an upper limit on H{\sc i} column density of $<$3.79$\times$10$^{20}$ cm$^{-2}$.  It has a flat spectrum at low frequencies ($\alpha_{\rm 150 MHz}^{\rm 1.4 GHz} \sim$0.12) 
 
 \emph{J114538.51+442021.9:} This source is a LERG at the redshift of 0.29974. A significant portion of the H{\sc i} spectrum of this source is also affected by the RFI at 1090 MHz and hence needs further checking.  However, from the velocities $-$200 to +400 km s$^{-1}$ relative to the optical systemic velocity we obtained a 3$\sigma$ upper limit on $N$(H{\sc i}) of 0.95$\times$10$^{20}$ cm$^{-2}$. The low-frequency spectral index for this source is flat ($\alpha_{\rm 150 MHz}^{\rm 1.4 GHz} \sim$0.46). This source is also identified among $\gamma$-ray emitting blazars and it could be a BL Lac object \citep{dbrusco2014ApJS..215...14D,penaherazao2021AJ....161..196P}. 
 
 \emph{J115712.38$-$032107.7:} 
 This radio  AGN is a LERG at the optical redshift of 0.08202 corresponding to an H{\sc i} observing frequency of 1312.7352 MHz. This source also has a flat low-frequency radio continuum spectrum ($\alpha_{\rm 150 MHz}^{\rm 1.4 GHz} \sim$0.29). We obtained a 3$\sigma$ $N${(H{\sc i})} upper limit of $\sim$1.86$\times$10$^{20}$ cm$^{-2}$. 

\emph{J143806.13+190954.9:} This source is a HERG at a redshift of 0.18854 (the corresponding H{\sc i} frequency is 1195.0845 MHz). It has a steep spectrum low-frequency spectral index ($\alpha_{\rm 150 MHz}^{\rm 1.4 GHz} \sim$0.76), and hence it could be a compact steep spectrum or extended radio source not resolved in the FIRST image. We did not detect H{\sc i} absorption towards this source and provide an upper limit on column density of 2.72$\times$10$^{20}$ cm$^{-2}$.
 
 \emph{J235400.91$-$003449.5:} This is a LERG at a redshift of 0.32593 showing inversion at the low-frequency radio continuum spectrum with no detection in the TGSS. We obtained an upper limit of 3.25$\times$10$^{20}$cm$^{-2}$ on its H{\sc i} column density.

\section{Discussion}
\subsection{H{\sc i} absorption detection rates}
\begin{figure}
    \centering
    \includegraphics[scale=0.55]{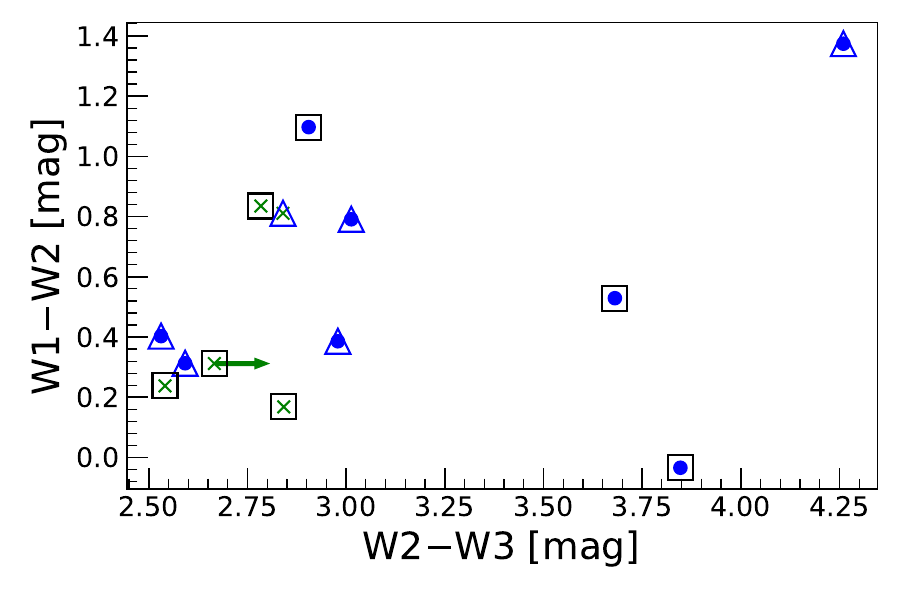}
    \caption{WISE W1$-$W2 vs W2$-$W3 colour-colour plot for the sample of 13 sources with H{\sc i} data not significantly affected with RFI. The symbols mean the same as in Fig.~\ref{fig:lumz}. The arrow means the lower limit.}
    \label{fig:wise_colors}
\end{figure}

In the literature, the dependence of H{\sc i } absorption detection rates on many factors such as the redshift, host galaxy star-formation history, AGN type, radio source structure, and radio and UV luminosities has been discussed. In this subsection, we explore the detection rates in our sample and discuss the effect of all these factors.
In our earlier work, for the sources at relatively lower redshift ($z<$0.25), we have found a strong dependence of H{\sc i} absorption detection rates on mid-infrared \textit{WISE} W2$-$W3 colours reflecting the star-formation history and dust-rich nature of the host galaxies \citep{chandola2017MNRAS.465..997C,chandola2020MNRAS.494.5161C}. Those with redder W2$-$W3$>$2 values were found to have higher detection rates ($\sim$50\%) than those with lower values ($\sim$10\%).  We selected our current sample to be dust and gas-rich with W2$-$W3 $>$ 2.5 occupying the region for luminous infrared galaxies (LIRGs) or ultra-luminous infrared galaxies(ULIRGs) or obscured AGN (see Fig.\ref{fig:wise_colors}) in the \textit{WISE} color-color diagram \citep{Wright2010AJ....140.1868W}. For the sources with no significant RFI effects, we find an overall H{\sc i} absorption detection rate of 8/13 (61.5$^{+30.4}_{-21.3}$\%) which is consistent with our earlier results.

\begin{figure}
    \centering
     \includegraphics[scale=0.55]{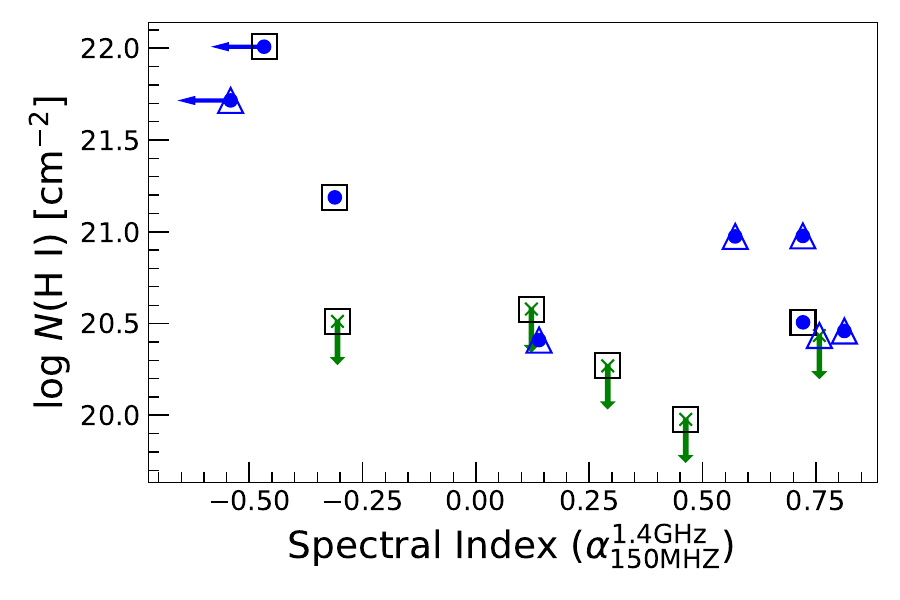}
     \includegraphics[scale=0.55]{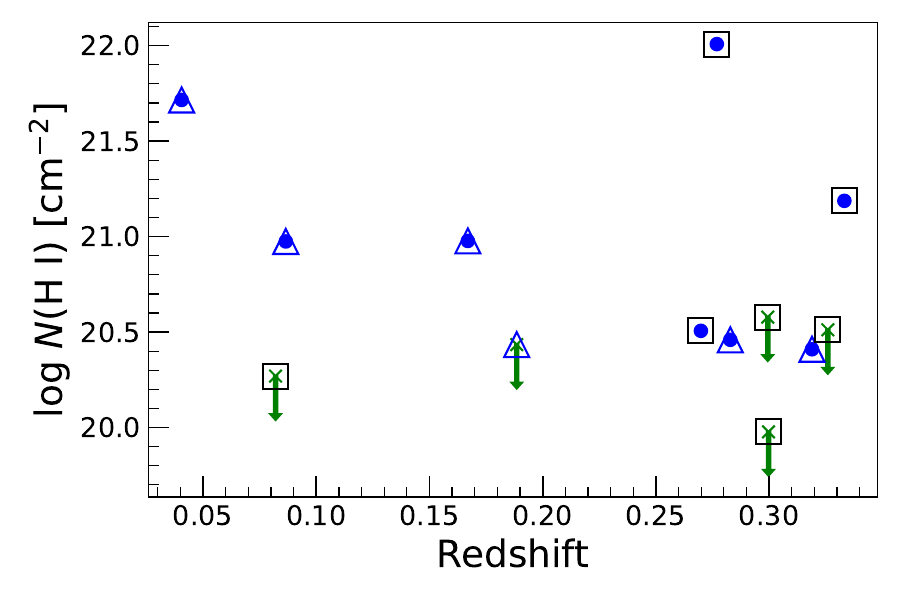}
    \includegraphics[scale=0.55]{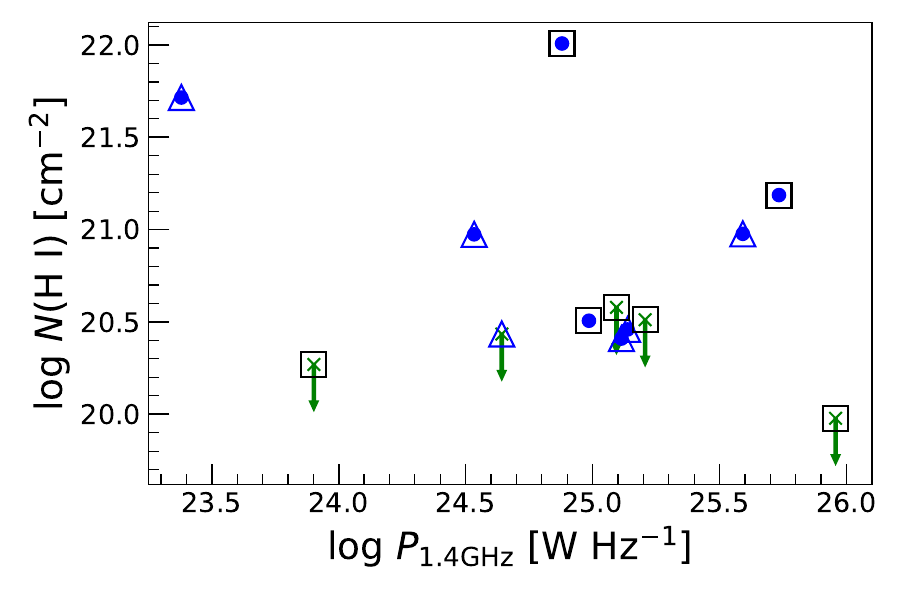}
    \caption{$N$(H{\sc i}) vs low-frequency spectral index ($\alpha_{\rm 150 MHz}^{\rm 1.4 GHz}$) in the top panel, $N$(H{\sc i}) vs redshift in the middle panel and $N$(H{\sc i}) vs log radio luminosity at 1.4 GHz in the bottom panel. The arrows mean upper limits. The rest of the symbols mean the same as in Fig.~\ref{fig:lumz}.}
    \label{fig:nhi_lum}
\end{figure}


We further consider the effect of low-frequency spectral indices ($\alpha_{\rm 150 MHz}^{\rm 1.4 GHz}$, $S_{\nu} \propto$ $\nu^{-\alpha}$) 
on the column densities. The low-frequency spectral indices ($\alpha_{\rm 150 MHz}^{\rm 1.4 GHz}$) can be used as a proxy of radio structures in the absence of high-resolution radio maps. Those showing turnover or inversion below 1.4 GHz are likely to be Peaked Spectrum sources.  Sources with a steep spectrum ($\alpha_{\rm 150 MHz}^{\rm 1.4 GHz}$ $>$0.5) are believed to have extended structures being compact steep spectrum (CSS) sources or larger in their projected linear sizes \citep{odea2021A&ARv..29....3O}. Most of the sources with H{\sc i} detection in our RFI-free sample (8/13) have either a flat ($\alpha_{\rm 150 MHz}^{\rm 1.4 GHz}$ $\sim$0.0 to 0.5) or inverted ($\alpha_{\rm 150 MHz}^{\rm 1.4 GHz}<$0) radio spectrum implying a compact nature. Most of the HERGs (4/6) in this sample are steep spectrum sources while LERGs (6/7) tend to have flat or inverted spectra. We find that there is an anti-correlation (Spearman's R-value: $-$0.71, p-value: 0.046) between H{\sc i} column densities and $\alpha_{\rm 150 MHz}^{\rm 1.4 GHz}$, with those with inverted spectrum having high H{\sc i} column densities (see Fig.~\ref{fig:nhi_lum} top panel). This is consistent with the scenario of compact radio sources tracing higher-density regions \citep{pihlstrom2003A&A...404..871P} and possibly confined due to high-density cold gas \citep{antao2012ApJ...760...77A}.  

Next, we consider the effect of central AGN accretion mode or strength. In our earlier paper, we found that the detection rates for LERGs and HERGs are similar at lower redshift ($z<$0.25)  if we have similar red mid-IR colours of host galaxies and compact radio structures \citep{chandola2020MNRAS.494.5161C}.
In this paper, we have expanded our work up to the redshift of 0.35.
Given the small sample, in our current survey, the detection rate for HERGs, 5/6 (83.3$^{+56.4}_{-36.0}$\%),   is similar within 1-$\sigma$ to the detection rate for LERGs, 3/7 (42.9$^{+41.7}_{-23.3}\%$)  which is consistent with our earlier results; namely if we consider LERGs and HERGs of similar redder WISE colours, detection rates are similar. 
Our sample has UV luminosities below 10$^{23}$ W Hz$^{-1}$, the cutoff to ionise or excite the hydrogen atoms to higher levels \citep{curran2024PASA...41....7C}, and hence there is no effect of this on detection rates.

Radio luminosities and redshift may also affect the detection rates with those with higher luminosity causing higher spin temperatures and evolution of gaseous environment with redshift. Of the 13 sources, 8 sources lie in the redshift range $z>$ 0.25 and have a detection rate of 5/8 (62.5$^{+42.3}_{-27.0} \%$) while those with $z<$0.25 have a detection rate of 3/5 (60$^{+58.4}_{-32.7}\%$). It shows statistically no significant difference  (see Fig.~\ref{fig:nhi_lum} middle panel).  Of the 8 sources at higher redshift six are LERGs, and three detections are LERGs and the other two are HERGs. At the lower redshift, 4/5 sources are HERGs and all three detections are HERGs. 
We also do not find any significant difference in H{\sc i} column densities with redshift for the sources with similar low-intermediate radio luminosities and \textit{WISE} colours, though it needs to be checked from a larger sample.

 Considering the full redshift range in our sample, we don't see any dependence of column density on radio luminosities. Our sample has sources with low-intermediate radio luminosities ($\lesssim$10$^{26}$ W/Hz) only (see Fig.~\ref{fig:nhi_lum} bottom panel). Now considering a limited redshift range of 0.25-0.4, there have been 67 sources (unaffected with RFI) studied in the literature but with only 10 detections reported before our study \citep{aditya2018MNRAS.473...59A, aditya2018MNRAS.481.1578A, curran2006MNRAS.371..431C, curran2011MNRAS.414L..26C, curran2017MNRAS.467.4514C, curran2019MNRAS.484.1182C, gupta2006MNRAS.373..972G,mahony2022MNRAS.509.1690M, murthy2021A&A...654A..94M,ostorero2017ApJ...849...34O, su2023ApJ...956L..28S, vermeulen2003A&A...404..861V,yan2016AJ....151...74Y,grasha2019ApJS..245....3G}.  \cite{murthy2021A&A...654A..94M} studied a sample of intermediate radio luminosity ($\sim$ 10$^{25.6}$-10$^{26.6}$ W Hz$^{-1}$ at 1.4 GHz) sources at $z \sim$ 0.25-0.4 and found a detection rate of 5/26 ($\sim$19$\%$). Detections in their sample are concentrated at lower redshifts with a detection rate of 5/15 (33.3$^{+22.5}_{-14.4}$\%) in the redshift range of 0.25-0.35.
As mentioned in Section \ref{sec:sec2}, sources at similar redshifts in our sample are of relatively low radio luminosities. While J145844.79+372021.5 is a common detection, four of our detections, J090410.36+024744.8, J122228.47+171437.3, J133242.53+134253.8 and J235515.92$-$011216.8 are among the lowest radio luminosity at the redshift range of 0.25-0.4. In this redshift range, earlier studies  by \cite{curran2006MNRAS.371..431C, curran2011MNRAS.414L..26C,curran2017MNRAS.467.4514C,curran2019MNRAS.484.1182C} have no detection from 15 sources. Similarly, \cite{aditya2018MNRAS.481.1578A,aditya2018MNRAS.473...59A} have a very low detection rate with only one detection towards J145844.79+372021.5 from 10 sources.  However, \cite{vermeulen2003A&A...404..861V} have a detection rate of 5/13 (38.5$^{+26.0}_{-16.6}$\%) towards radio sources of higher radio luminosities ($\gtrsim$10$^{26}$ W Hz$^{-1}$). The different detection rates in these studies are possibly due to different sample selection criteria and optical depth sensitivities of observations. We find our detection rates (5/8, 62.5$^{+42.3}_{-27.0}\%$ ) consistent with those of \cite{vermeulen2003A&A...404..861V} within 1$\sigma$ implying no significant effect of radio luminosities on detection rates. However, our and \cite{vermeulen2003A&A...404..861V} sample sizes are small in this limited redshift range and need to be probed from a larger sample.

\begin{figure}
    \centering
     \includegraphics[scale=0.55]{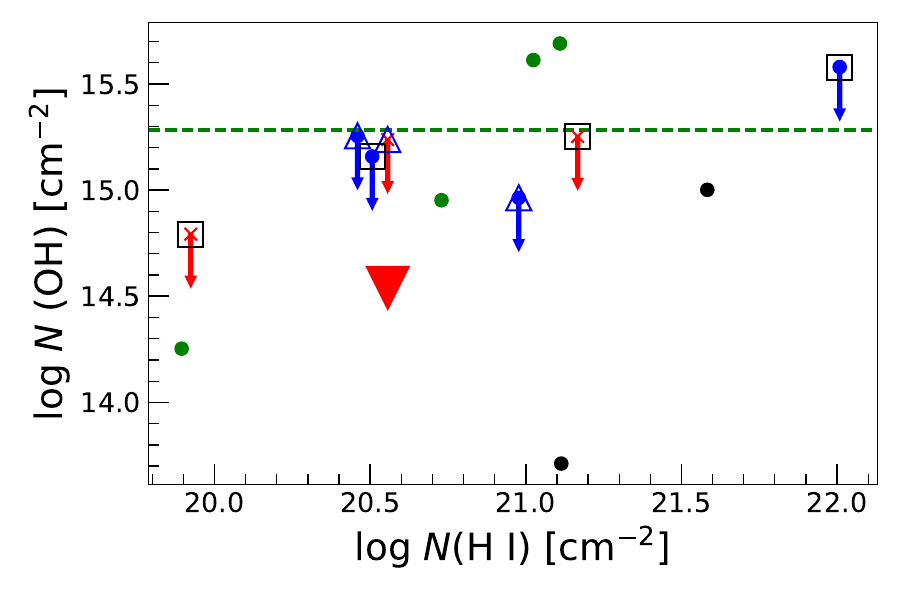}
     \includegraphics[scale=0.55]{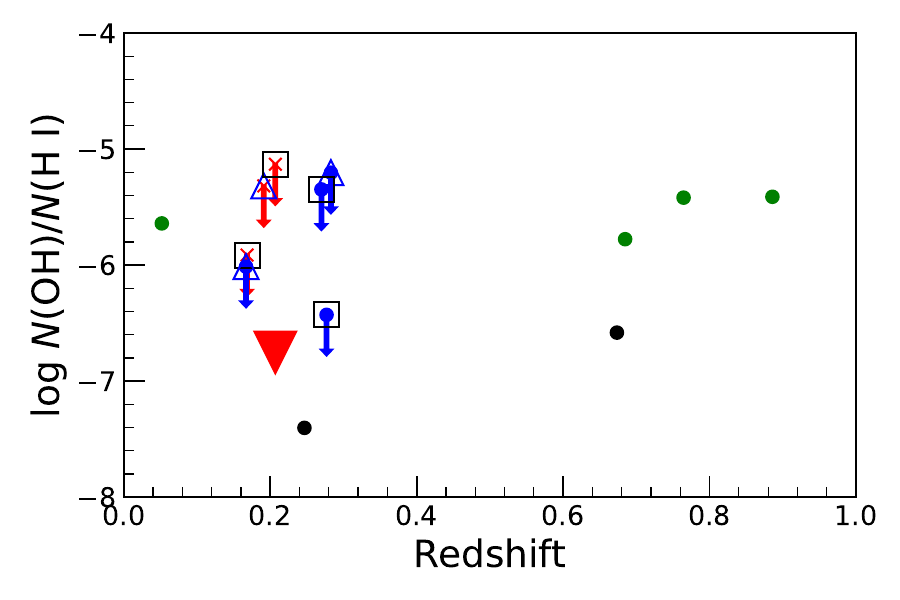}
    \caption{Top panel: $N$(OH) vs $N$(H{\sc i}), Bottom panel: $N$(OH)/$N$(H{\sc i}) vs redshift.
    Green and black-filled circles represent previous OH absorption detections from literature towards intervening and associated  H{\sc i} absorption systems \citep{ kanekar2002A&A...381L..73K,kanekar2003A&A...412L..29K,kanekar2012ApJ...746L..16K,gupta2018ApJ...860L..22G} respectively which have been taken from \cite{zheng2020MNRAS.499.3085Z} compilation. In the top panel, green dashed horizontal line marks the median value of $N$(OH) for four intervening  OH absorbers in the literature. The red downward triangles represent the upper limit values from stacking at their median $N$(H{\sc i}) and redshift values in the top and 
    bottom panels respectively. The sources affected with RFI in our observation but reported with H{\sc i} detection in literature are shown with red crosses.  The arrows mean upper limits. Other symbols mean the same as in Fig.~\ref{fig:lumz}. To estimate $N$(OH), we have used $T_{ex}=$10 K and $f_{\rm c}$(OH)=1 for all sources in these plots. Similarly for $N$(H{\sc i}), we have used $T_{\rm s}=$100 K and $f_{\rm c}$(H{\sc i})=1 for all sources.}
    \label{fig:noh_nhi}
\end{figure}

\subsection{H{\sc i} absorption kinematics}
H{\sc i} absorption kinematics combined with the radio  AGN properties such as morphology and radio power have been used to interpret the nature of absorbing medium in the host galaxy of radio  AGN in the literature. The interpretations of H{\sc i} absorption profiles are largely based on the profile parameters such as line width, its position relative to optical emission lines, the depth of the profile and the number of components. The absorption profiles near the optical emission lines are interpreted as rotating disks (e.g. 3C 84, \citealt[][]{morganti2023A&A...678A..42M}; 4C 31.04, \citealt[][]{murthy2024arXiv240517389M}) which could be circumnuclear torus or gas in the host galaxy. If the profile is redshifted w.r.t optical lines, it could also be due to infalling gas (e.g. NGC 315, \citealt[][]{morganti2009A&A...505..559M}; B2352+495, \citealt[][]{araya2010AJ....139...17A}; 4C 31.04, \citealt[][]{struve2012A&A...546A..22S}).  Broad, shallow and blueshifted absorption lines are usually interpreted as outflowing gas due to AGN or stellar feedback effects (e.g. IC 5063, \citealt[][]{morganti1998AJ....115..915M}; 3C 293, \citealt[][]{morganti2003ApJ...593L..69M}; 4C 12.50, \citealt[][]{morganti2013Sci...341.1082M}). It was found in earlier works that the compact sources show more incidences of blue-shifted profiles suggesting outflowing gas \citep{chandola2013MNRAS.429.2380C, gereb2015A&A...575A..44G,maccagni2017A&A...604A..43M}. Also, the sources with higher radio power were found to have larger blueshifted and wider profiles \citep{chandola2011MNRAS.418.1787C, gereb2015A&A...575A..44G, maccagni2017A&A...604A..43M, chandola2017MNRAS.465..997C, chandola2020MNRAS.494.5161C}. Due to the low-intermediate radio luminosity, the sources in our sample do not show a large shift w.r.t optical redshift, except for J133242.53+134253.8 which is blueshifted by $\sim$300 km s$^{-1}$. Also, except for the sources, J090410.36+024744.8, J133242.53+134253.8  and J135223.46$-$015648.4, all other absorption lines in our sample are narrow lines with their FWHM $<$ 100 km s$^{-1}$ indicating a settled H{\sc i} gas in the galaxy disk. Possibly, most of the radio sources in our sample are either too compact and/or less inclined towards the disk to have any significant jet-ISM interaction as predicted by simulations \citep{mukherjee2018MNRAS.479.5544M,mukherjee2018MNRAS.476...80M}.
For J090410.36+024744.8, multiple components of the Gaussian could be fitted reflecting multiple absorbers in the interstellar medium, although all components are within $\pm$100 km s$^{-1}$. In J135223.46$-$015648.4 the narrow, deep component redshifted by $\sim$155 km s$^{-1}$ could be infalling gas while the broader component could be a cloud closer to the central region similar to the cases of NGC 315 and 4C 31.04 \citep{morganti2009A&A...505..559M, struve2012A&A...546A..22S}. Another possibility could be that the narrow profile is from H{\sc i} disk but the orientation of the background radio source is such that the gas in front appears redshifted relative to optical systemic velocity. In that scenario, the broader component may be due to jet-cloud interaction causing the outflow of cold gas similar to 3C 293 \citep{morganti2003ApJ...593L..69M}. In the case of J133242.53+134253.8, the broad, shallow and blue-shifted absorption profile could be due to jet-cloud interaction or unsettled gas from an earlier merger episode.  It would require a detailed parsec scale study to have further insights into detailed kinematics for these sources.

\subsection{OH absorption and abundance}
In Fig.~\ref{fig:noh_nhi}, we explore the abundance of OH compared to atomic H{\sc i} gas. In an earlier paper, \citet{zheng2020MNRAS.499.3085Z} found that [OH]/[H{\sc i}] increased with redshift and for the associated H{\sc i} absorbers the ratio has a lesser value compared to intervening absorbers in literature. However, in our analysis, we find that sources in our sample with H{\sc i} absorption have the upper limits on $N$(OH) being higher. Sources in our sample also have a higher average value of $N$(OH)/$N$(H{\sc i}) upper limits compared to \citet{zheng2020MNRAS.499.3085Z} for a similar redshift range of $z<$0.35.  We have similar assumptions on excitation temperature and covering factor, hence it is the higher upper limits on integrated optical depths that play a role. Since the spectral \textit{rms} on optical depths are similar, it is because they are integrating within FWHM and we are integrating within a Gaussian FWZI (nearly 2.547 $\times$FWHM) and they have a better spectral resolution of $\lesssim$ 2 km s$^{-1}$ which results in a lower upper limit in their case.  Also, sources in our sample cover a wider range of H{\sc i} column densities with most of the sources (5/7) below 10$^{21}$ cm$^{-2}$ while \cite{zheng2020MNRAS.499.3085Z} have sources above 10$^{21}$ cm$^{-2}$. Due to this, we have a higher upper limit on [OH]/[H{\sc i}] for similar redshift.

Although even after stacking upper limits on OH column densities are lower for \cite{zheng2020MNRAS.499.3085Z} due to the above reasons, the optical depth \textit{rms} obtained by us are slightly better suggesting these systems may have  OH below our detection limit. Stacking results confirm the lower 3$\sigma$ upper limits on $N$(OH) $\sim$3.47$\times$10$^{14}$ cm$^{-2}$ for 7 associated H{\sc i} absorbers compared to the median value on $N$(OH)$\sim$1.92$\times$10$^{15}$ cm$^{-2}$ for 4 intervening H{\sc i} and OH absorption systems of similar H{\sc i} column densities from literature. It also confirms the lower 3$\sigma$ upper limits on [OH]/[H{\sc i}] $\sim$1.78 $\times$10$^{-7}$ in low-intermediate redshift associated H{\sc i} systems compared to the value 2.62$\times$10$^{-7}$ for only higher redshift ($z\sim$0.673) associated H{\sc i} and OH absorber, B3 1504+377 \citep{kanekar2002A&A...381L..73K}.

\section{Conclusions}
We have the following conclusions from our study, although our observations towards many sources (26 at H{\sc i} frequencies and 7 at OH frequencies) were affected by RFI.
\begin{enumerate}

    \item 
    We report the H{\sc i} absorption detection towards 8 sources from 13 sources with good data. Of these eight, five absorbers are discoveries and four detections are of the lowest radio luminosity sources in the redshift range 0.25-0.4. The H{\sc i} column densities for assumed spin temperature of 100 K and covering factor of one are in the range $\sim$2.57$\times$10$^{20}$ cm$^{-2}$  to 101.92 $\times$10$^{20}$ cm$^{-2}$ with a median value 9.46 $\times$10$^{20}$ cm$^{-2}$. For the H{\sc i} non-detections, median 3$\sigma$ upper limit on H{\sc i} column densities is 2.72 $\times$10$^{20}$ cm$^{-2}$. Except for the sources, J090410.36+024744.8, J133242.53+134253.8  and J135223.46$-$015648.4, all other absorption lines in our sample are narrow lines with their FWHM $<$ 100 km s$^{-1}$. 
  
    \item We find that the detection rates are primarily affected by the red mid-infrared color reflecting the dust and gas-rich nature of the host galaxy and the compactness of the radio sources. We do not find any significant dependence of the detection rates on either radio luminosity or redshift. 
    \item We find that the H{\sc i} column densities are anti-correlated to low-frequency spectral indices ($\alpha_{\rm 150 MHz}^{\rm 1.4 GHz}$, $S_{\nu} \propto\nu^{-\alpha}$). 
    Those with an inverted continuum spectrum have higher column densities. 
    \item  For similar mid-IR colors, detection rate for HERGs is similar within 1$\sigma$ to that of LERGs which is consistent with our previous results for lower redshift objects. 
    
    \item We do not detect OH absorption in any of the 23 sources with good data at OH frequencies. However, from stacking the spectra, we obtain a stringent upper limit on OH column densities to be 3.47$\times$10$^{14}$ $T_{\rm ex}$/10 K $\times$1/$f_{\rm c}$ cm$^{-2}$ and $N$(OH)/$N$({H\sc i}) ratio $=$1.78 $\times$10$^{-7}$ for sources with H{\sc i} absorption. Upon stacking, we find OH column density 3$\sigma$ upper limit for associated absorbing systems is lower than intervening ones and  [OH]/[H{\sc i}] upper limit for low-intermediate redshift systems is lower than high redshift absorbers. 
    \item Large ongoing H{\sc i} and OH absorption surveys with SKA pathfinders will be useful to probe these findings further using larger samples.
    
\end{enumerate}

\begin{acknowledgements}
We thank an anonymous reviewer for the useful comments and suggestions which helped to improve the paper significantly.
This work has used the data from the Five-hundred-meter Aperture Spherical radio Telescope (FAST).  FAST is a Chinese national mega-science facility, operated by the National Astronomical Observatories of the Chinese Academy of Sciences (NAOC). 
YC thanks the Center for Astronomical Mega-Science, Chinese Academy of Sciences, for the FAST distinguished young researcher fellowship (19-FAST-02). YC also acknowledges the support from the National Natural Science Foundation of China (NSFC) under grant No. 12050410259 and the Ministry of Science and Technology (MOST) of China grant no. QNJ2021061003L. YZM acknowledges the support from the National Research Foundation of South Africa with Grant No. 150580, No. 159044, No. CHN22111069370 and No. ERC23040389081. ZZ is supported by NSFC grant No. 11988101, 12373012, U1931110, 12041302 and the Young Researcher Grant of Institutional Center for Shared Technologies and Facilities of National Astronomical Observatories, Chinese Academy of Sciences. CWT was supported by a grant from the NSFC (No. 12041302). DT acknowledges financial support from the XJTLU Research Development Fund (RDF) grant with number RDF-22-02-068. HP acknowledges support from a UKRI Frontiers Research Grant [EP/X026639/1], which was selected by the ERC. 

This publication makes use of data products from the Wide-field Infrared Survey Explorer, which is a joint project of the University of California, Los Angeles, and the Jet Propulsion Laboratory/California Institute of Technology, and NEOWISE, which is a project of the Jet Propulsion Laboratory/California Institute of Technology. WISE and NEOWISE are funded by the National Aeronautics and Space Administration. Funding for the Sloan Digital Sky Survey (SDSS) has been provided by the Alfred P. Sloan Foundation, the Participating Institutions, the National Aeronautics and Space Administration, the National Science Foundation, the U.S. Department of Energy, the Japanese Monbukagakusho, and the Max Planck Society. The SDSS Web site is http://www.sdss.org/. The SDSS is managed by the Astrophysical Research Consortium (ARC) for the Participating Institutions. The Participating Institutions are The University of Chicago, Fermilab, the Institute for Advanced Study, the Japan Participation Group, The Johns Hopkins University, Los Alamos National Laboratory, the Max-Planck-Institute for Astronomy (MPIA), the Max-Planck-Institute for Astrophysics (MPA), New Mexico State University, University of Pittsburgh, Princeton University, the United States Naval Observatory, and the University of Washington.
\end{acknowledgements}

\facility{FAST \citep{Nan2011IJMPD..20..989N}}
\software{Python, Numpy\citep{harris2020array}, Matplotlib\citep{Hunter:2007}, Astropy\citep{astropy:2013, astropy:2018, astropy:2022}, TOPCAT\citep{taylor2005ASPC..347...29T}}

\bibliographystyle{aasjournal}


\end{document}